\documentclass[twocolumn,showpacs,preprintnumbers,amsmath,amssymb]{revtex4}
\usepackage{bm}
\usepackage{mathrsfs}
\usepackage{subfigure}
\usepackage{graphicx}
\newcommand{\ua}{$\mathrm{U}_{\rm A}(1)$ }

\begin{document}
\preprint{CYCU-HEP-11-20}
\title{
Nonet meson properties in Nambu--Jona-Lasinio model with dimensional\\
regularization at finite temperature and chemical potential
}
\author{T. Inagaki}
\affiliation{
Information Media Center, Hiroshima University,
Higashi-Hiroshima, Hiroshima 739-8521, Japan
}
\author{D. Kimura}
\affiliation{
Faculty of Education, Hiroshima University,
Higashi-Hiroshima, Hiroshima 739-8524, Japan
}
\author{H. Kohyama}
\affiliation{
Department of Physics, Chung-Yuan Christian University,
Chung-Li 32023, Taiwan
}
\author{A. Kvinikhidze}
\affiliation{
A. Razmadze Mathematical Institute of Georgian Academy of Sciences,\\
M. Alexidze Str. 1, 380093 Tbilisi, Georgia
}

\date{\today}
\begin{abstract}
The nonet meson properties are studied in the Nambu--Jona-Lasinio model 
at finite temperature and chemical potential using dimensional regularization. 
This study leads to the reasonable description which is mainly
similar to one obtained in the model with the cutoff regularization. 
However, remarkable differences between the two regularizations are observed
in the behavior of the chiral phase transition at finite chemical potential.
\end{abstract}
\maketitle

\section{INTRODUCTION}
Mesons are composed of quarks, 
interacting with each other through the exchange of gluons.
Under the low temperature and density conditions, quarks are confined 
in hadrons and therefore are not observed as free particles, which makes 
their investigation challenging. The phenomenon of
confinement is closely related to the dynamical chiral symmetry
breaking. While the chiral symmetry broken in vacuum is expected to be
restored at high temperature and density, it is also expected that the
properties of mesons are affected by this chiral phase transition.
Thus the investigation of meson properties at finite temperature
and density is an important part of hadron physics.

Such investigation unfortunately cannot be pursued directly in the
framework  of quantum chromodynamics (QCD), the ultimate theory of
strong interaction, due to its large coupling constant at low energy
scale and thereby the necessity of dealing non-perturbatively with
its notoriously complicated structure. The investigations are usually
performed by using the effective models which share the same symmetry
properties with QCD such as the Nambu--Jona-Lasinio (NJL) model~%
\cite{NJL}, the linear sigma model~\cite{LsM}, or by using the
discretized version of QCD, i.e., the lattice QCD~\cite{Wilson:1974sk}.

We employ, in this paper, the NJL model to study the nonet meson
properties (for reviews on the model, see e.g.,~\cite%
{Vogl:1991qt,Klevansky:1992qe,Hatsuda:1994pi,Buballa:2005rept}).
It is not renormalizable therefore one has to introduce the
regularization procedure to handle the divergences appearing in
loop integrals. There are several regularization schemes: the three-
and four-momentum cutoff regularization, the dimensional
regularization, the Pauli-Villars methods, and so on. We shall use
the dimensional regularization (DR) in this paper. In the DR
method the divergent fermion loop integrals which are
the functions of the space-time dimensions are regularized by applying
an analytic continuation in the dimensions variable. Many works
are devoted to the studies based on it (see e.g.,~\cite%
{Krewald:1991tz,Inagaki:1994ec,Jafarov:2004jw,Inagaki:2007dq,%
Fujihara:2008ae}).

We believe that a regularization procedure is an important 
{\it dynamical} part of the NJL model, it gives interaction between 
quarks the size and shapes it, which otherwise is point-like in the 
traditionally used leading order approximation of the $1/N_c$ expansion. 
That is why it is important to chose a proper regularization. Dimensional 
and cutoff regularizations can be considered as corresponding to the same 
size but different shape of the interaction. The fact that some physical 
results depend on the regularization way implies that the corresponding 
physics is sensitive (probes) not only to the size of the interaction 
but to its shape as well. If so the regularization dependent properties 
would be determined by higher energy (shorter distances) than those which 
do not depend on the regularization way.

The purpose of this paper is to study the nonet meson properties such
as masses and decay constants in the three-flavor NJL model with the
DR at finite temperature $T$ and chemical
potential $\mu$. To this end one has to firstly determine the model
parameters so that the model describes the physical ingredients
properly. This has been done in the paper~\cite{Inagaki:2010nb},
where the present authors have found that the various parameter sets
reproduce the meson properties nicely at $T=0$ and $\mu=0$. As our
goal is to study meson properties at finite $T$ and $\mu$, the paper
is the straight forward extension of the previous work~%
\cite{Inagaki:2010nb}.

This paper is organized as follows: In Sec.~\ref{njl_model}, the three-%
flavor NJL model and its parameters are introduced. In
Sec.~\ref{constituent_mass} we discuss the constituent quark masses and chiral
condensates which are derived through solving the gap equations.
The results on the meson properties and topological susceptibility
are shown in Sec.~\ref{meson} and \ref{topo}. Sec.~\ref{critical} is
devoted to the discussion of the critical temperature.
Summary and discussions are given in Sec.~\ref{conclusion}.
The required calculations of the chiral condensates and meson
properties are aligned in App.~\ref{app_chiral} and \ref{app_meson}.
The fitted parameters are shown in App.~\ref{app_para}.

\section{NJL model with \ua anomaly}
\label{njl_model}
\subsection{The model}
The Lagrangian of our three-flavor NJL model is 
\begin{equation}
  \mathcal{L}_{\mathrm{NJL}}
   = \sum_{i,j} \bar{q}_i\left( i \partial\!\!\!/
       - \hat{m}\right)_{ij}q_j + \mathcal{L}_4 + \mathcal{L}_6 ,
  \label{LNJL}
\end{equation}
where
\begin{align}
  \mathcal{L}_4
  &= G \sum_{a=0}^8 \biggl[
       \Bigl( \sum_{i,j} \bar{q}_i\lambda_a q_j\Bigr)^2
     + \Bigl( \sum_{i,j}\bar{q}_i\,i \gamma_5 \lambda_a q_j \Bigr)^2
     \biggr] ,
  \label{L_4} \\
  \mathcal{L}_6
  &= -K \left[ \det\bar{q}_i (1-\gamma_5) q_j 
     +\text{h.c.\ } \right] .
  \label{L_6}
\end{align}
Here the subscripts $i,j$ represent the flavor indices, $\hat{m}$ is
the current quark mass matrix diag$(m_u,m_d,m_s)$, and $\lambda_a$
are the Gell-Mann matrices in flavor space with
$\lambda_0=\sqrt{2/3}\cdot{\bf 1}$. $\mathcal{L}_4$ and $\mathcal{L}_6$
are the four- and six-fermion interactions with the effective coupling
constants $G$ and $K$. $\mathcal{L}_6$ is introduced to break the \ua
symmetry which is not realized in the real world. The six-quark vertex
in the determinant form was firstly discussed by Kobayashi and Maskawa~%
\cite{Kobayashi:1970ji} and later derived by 't Hooft as an
instanton-induced quark interaction~\cite{'tHooft:1976fv}, so this
vertex is called Kobayashi-Maskawa-'t Hooft term.

To study the thermal system we evaluate the thermodynamic potential
$\Omega = -\ln Z / V\beta$, with the partition function $Z$, the
volume of the system $V$ and the inverse temperature $\beta(\equiv 1/T)$.
Here we use the imaginary time formalism to treat the system at
finite $T$ and $\mu$.
Applying the mean-field approximation in the DR
scheme, we obtain the following thermodynamic potential;
\begin{align}
  \Omega =
  &2G(\phi_u^2+\phi_d^2+\phi_s^2) -4K \phi_u \phi_d \phi_s \nonumber\\
  &-\frac{2^{D/2}N_c}{2} \int \!\! \frac{d^{D-1}p}{(2\pi)^{D-1}}
             \bigl[ E_u + E_d + E_s \bigr]  \nonumber \\
  &-\frac{2^{D/2}N_c}{2} T \int \!\! \frac{d^{D-1}p}{(2\pi)^{D-1}}
             \sum_{i,\,\pm} \ln \Bigl[1 + e^{-\beta E_i^{\pm}} \Bigr],
  \label{thermo}
\end{align}
where $\phi_i\equiv \langle \bar{i}i \rangle$ represent the chiral
condensates and $N_c(=3)$ is the number of colors.
$E_i=\sqrt{p^2+m_i^{*\,2}}$ is the energy of the quasi-particle, and
$E_i^{\pm}=E_i\pm \mu$, with the constituent quark masses $m_i^*$ and
a chemical potential $\mu$. In this scheme we
regularize the divergent integrals by performing the analytic
continuation of the space-time dimension $D$ to a value less than
four, then the integral can be written as
\begin{align}
  \int \!\frac{d^{D-1}p}{(2\pi)^{D-1}} 
  =
  \frac{2\,(4\pi)^{-(D-1)/2}}{\Gamma[(D-1)/2]} M_0^{4-D}
    \int_0^{\infty} \!\!dp \,p^{D-2} ,
\end{align}
where $M_0$ is a renormalization scale.
One can switch to cutoff regularization through the replacement
\begin{align}
  \int \!\frac{d^{D-1}p}{(2\pi)^{D-1}} 
  \rightarrow
  \frac{1}{2\pi^2} \int_0^{\Lambda} \!\!dp \,p^{2}.
\end{align}

The chiral condensates $\phi_i$ which are the order parameters of
the model are evaluated through solving the gap equations,
$\partial \Omega / \partial \phi_i=0$. After some algebra, one
arrive at the following self-consistent equations,
\begin{align}
  m_i^{*} = m_i - 4 G \phi_i + 2 K \phi_j \phi_k,\, (i \neq j \neq k)
  \label{gap}
\end{align}
where
\begin{align}
  &\phi_i 
          = \frac1\beta \sum_{-\infty}^\infty
  \int \!\! \frac{d^{D-1}p}{(2\pi)^{D-1}}\,{\rm tr} S^i(p), 
  \label{trace} \\
  &S^i(p) = \frac1{{\boldsymbol p}\cdot {\boldsymbol \gamma}
   -(\omega_n -i\mu)\gamma_4 +m_i^* -i\epsilon}, \nonumber
\end{align}
with $\omega_n=(\pi/\beta)(2n+1)$, ($n=0,\pm1,\pm2,\cdots$).
The tr stands for the trace in spinor and color indices. The explicit
form of the integral is presented in the App.~\ref{app_chiral}.

To evaluate meson properties, one needs to carry out further
calculations as discussed in~\cite{Inagaki:2010nb} where only the
case of $T,\mu=0$ is presented. We carry out the corresponding
calculations for the case of finite $T$ and $\mu$ in the App.~%
\ref{app_meson}.

\subsection{Model parameters}
In the DR, the three-flavor NJL model has seven
parameters:
\begin{quote}
current quark masses  \,\,$m_u$, $m_d$, $m_s$\\
four-point coupling   \,\,$G$ \\
six-point coupling    \,\,$K$ \\
dimensions            \,\,$D$ \\
renormalization scale     $M_0$.
\end{quote}
Since the mass difference between $m_u$ and $m_d$ is small as
compared to one between $m_s$ and $m_u$ (or $m_d$), we take the
isospin limit, $m_d=m_u$, and test several values,
$m_u=3,\,4,\,5,\,5.5,\,6$MeV. To fix the other parameters, we usually
employ the following physical observables~\cite{PDG}:
\begin{align}
  \begin{array}{cc}
    m_{\pi}=138{\rm MeV}, & f_{\pi}=92 {\rm MeV}, \\ 
    m_{\mathrm K}=495 {\rm MeV}, & m_{\eta^{\prime}}=958 {\rm MeV}.
  \end{array}
\label{input}
\end{align}
One more physical quantity is needed to fix all seven above listed
parameters. Possible candidates are the alternative meson properties
such as the $\eta$ meson mass, $m_{\eta}$, topological susceptibility,
$\chi$, kaon decay constant, $f_{\mathrm K}$, etc. We choose the
following three cases:
\begin{quote}
 Case $m_{\eta}$\ \, $m_{\pi},\,m_{\mathrm K},\,f_{\pi},\,m_{\eta^{\prime}}$,
$m_{\eta}=548$MeV\\
\hspace{-10mm}
Case $\chi_{170}$ $m_{\pi},\,m_{\mathrm K},\,f_{\pi},\,m_{\eta^{\prime}}$,
$\chi^{1/4}=170$MeV\\
Case $\chi_{179}$ $m_{\pi},\,m_{\mathrm K},\,f_{\pi},\,m_{\eta^{\prime}}$,
$\chi^{1/4}=179$MeV
\end{quote}
As mentioned in the introduction, the parameters are fixed
at $T,\mu=0$ in~\cite{Inagaki:2010nb}; we show the complete sets
of them in the App.~\ref{app_para}.  We do not consider the $T$ and
$\mu$ dependence of these parameters in this paper. In App.~%
\ref{app_para}, we also show the sets in the three-momentum sharp
cutoff case to make comparison of the two different
regularization schemes. In the cutoff regularization the six parameters,
$m_u,m_d(=m_u),m_s,G,K$ and cutoff scale $\Lambda$, are fixed by using
the four physical observables (\ref{input}) and choosing several values of
$m_u$.

\section{Constituent quark masses}
\label{constituent_mass}

Before going through the meson properties, let us discuss the results
for the constituent quark masses $m_i^*$. This is important because the
constituent quark masses determine the scale of the system, and they
play a crucial role in the studies of the transition temperature and
chemical potential.

\subsection{$m^*$ in the DR scheme}
\begin{figure}[!h]
  \begin{center}
    \includegraphics[width=3.2in,keepaspectratio]{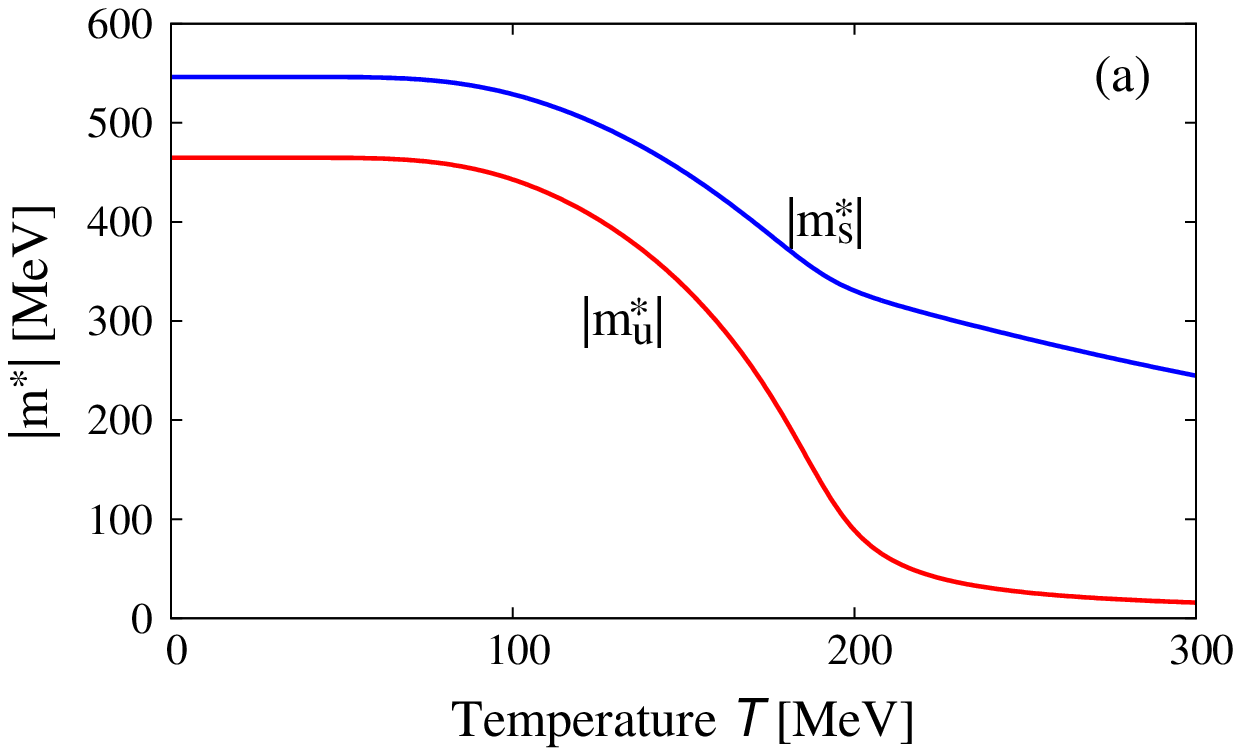}
    \includegraphics[width=3.2in,keepaspectratio]{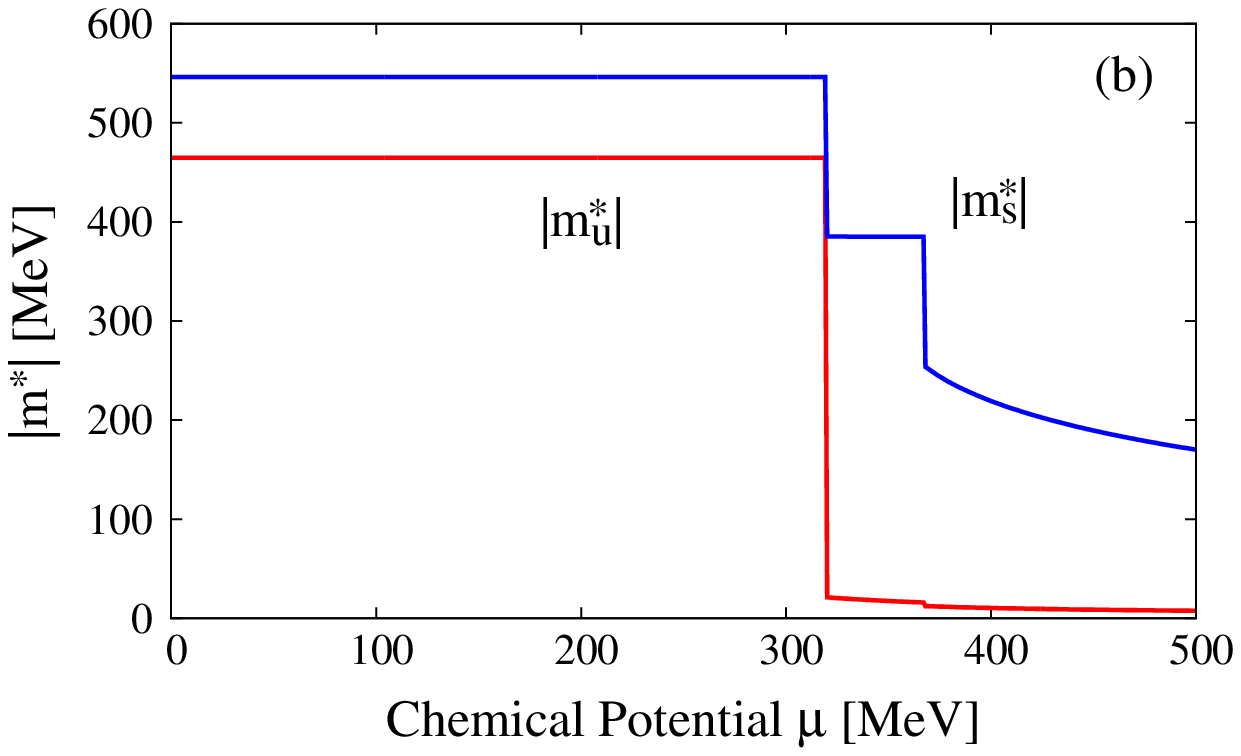}
  \end{center}
  \vspace{-0.3cm}
  \caption{$m_i^*$ in the Case $m_{\eta}^{\rm LD}$ with $m_u=3$MeV.
           (a) finite $T$ and $\mu=0$.
           (b) finite $\mu$ and $T=10$MeV.}
  \label{M_etaLow}
\end{figure}
In Fig.~\ref{M_etaLow}, we display the numerical results for the
constituent quark masses which are obtained by solving the gap
equations Eq.(\ref{gap}) as explained in the previous section. We
note that the absolute values of $m_u^*$ and $m_s^*$ gradually
decrease with increasing temperature, while at low temperature they
do not depend on the chemical potential up to its critical
value where they suddenly fall.
This is a clear signal of the first order phase transition.

\begin{figure}
  \begin{center}
    \includegraphics[width=3.2in,keepaspectratio]{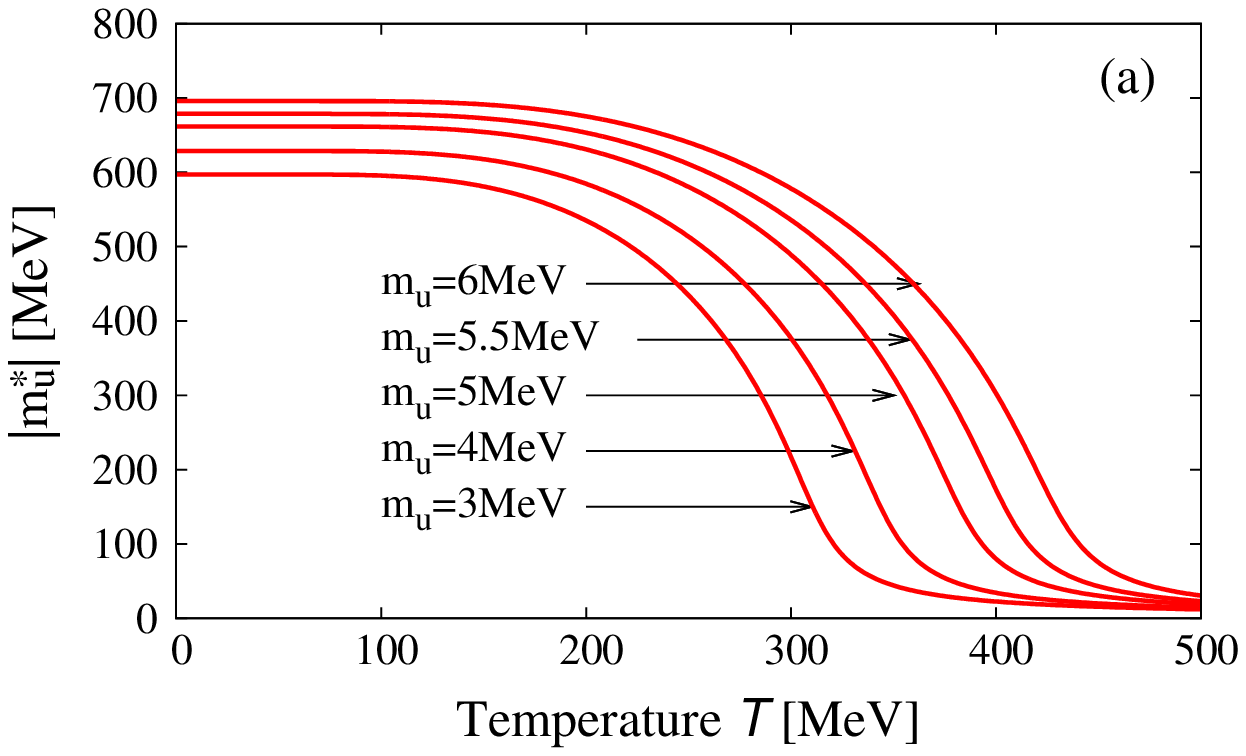}
    \includegraphics[width=3.2in,keepaspectratio]{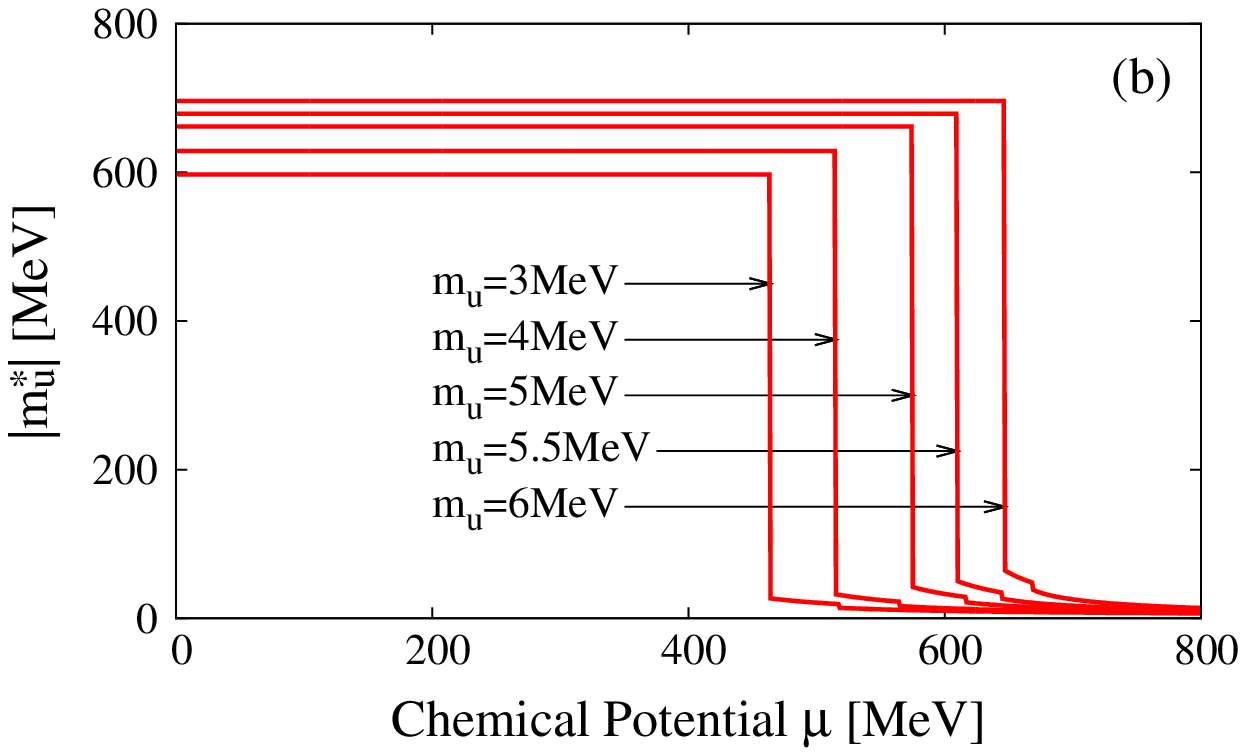}
  \end{center}
  \vspace{-0.3cm}
  \caption{$m_u^*$ in the Case $m_{\eta}$ with various $m_u$. 
           (a) finite $T$ and $\mu=0$.
           (b) finite $\mu$ and $T=10$MeV.}
  \label{Mu_eta}
\end{figure}
\begin{figure}
  \begin{center}
    \includegraphics[width=3.2in,keepaspectratio]{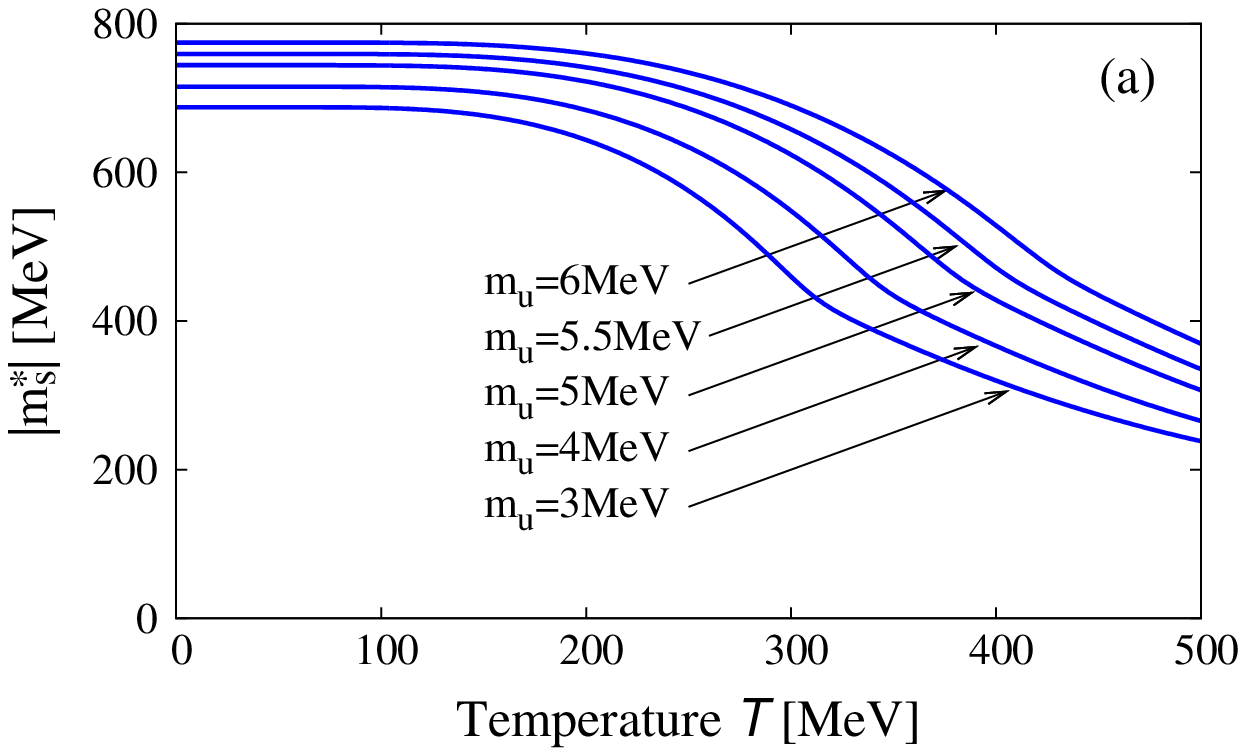}
    \includegraphics[width=3.2in,keepaspectratio]{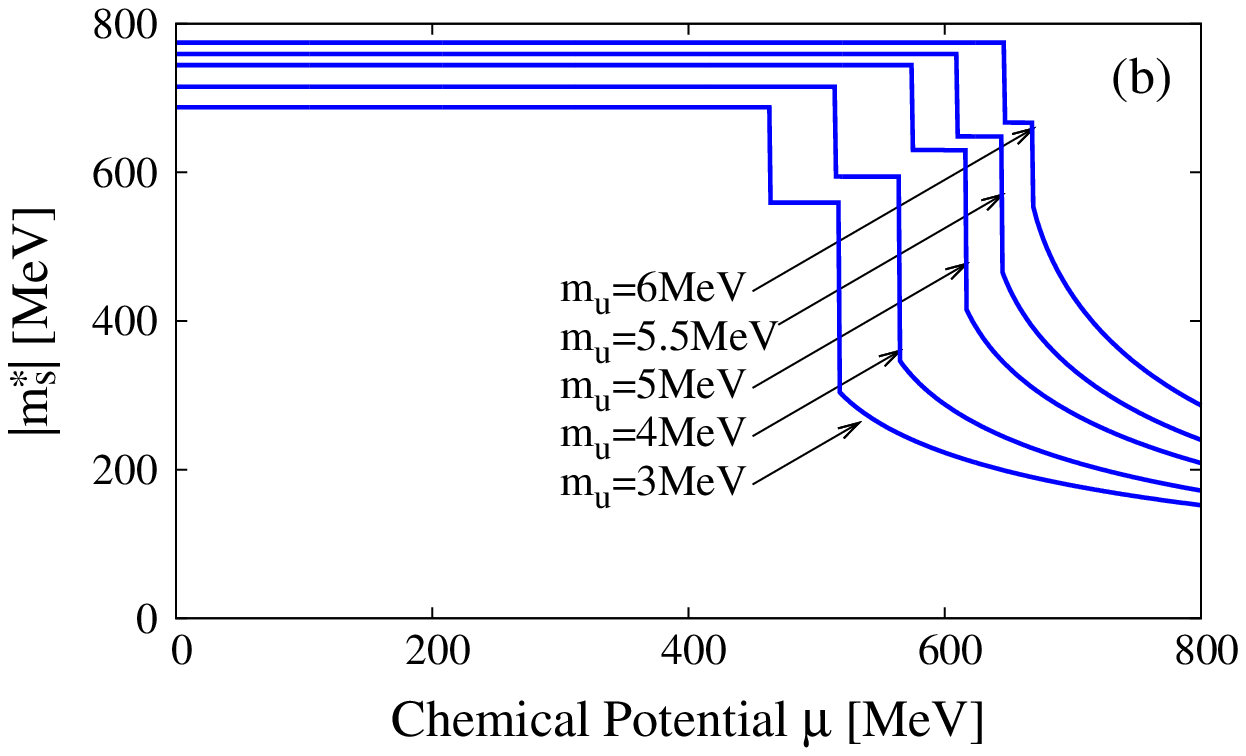}
  \end{center}
  \vspace{-0.3cm}
  \caption{$m_s^*$ in the Case $m_{\eta}$ with various $m_u$. 
           (a) finite $T$ and $\mu=0$.
           (b) finite $\mu$ and $T=10$MeV.}
  \label{Ms_eta}
\end{figure}
Similar pictures correspond to the different values of $m_u$ as it is
shown in Fig.~\ref{Mu_eta} where the dependence of $m_u^*$ on $T$ and
$\mu$ is given in the Case $m_{\eta}$ for fixed values of $m_u=3$, $4$,
$5$, $5.5$ and $6$MeV. These pictures again indicate the crossover
transition for low $\mu$ and  the first order phase transition for
low $T$. Thus the behavior of the first order and crossover
transitions holds for the different values of $m_u$ in the Case
$m_{\eta}$.

In Fig.~\ref{Ms_eta} there are shown $m_s^*$ for different values of $m_u$
as the functions of $T$ or $\mu$ in the Case $m_\eta$. 
According Eq.~(\ref{gap}), due to the large contribution of $m_s$, 
the absolute value of $m_s^*$ is larger than that of $m_u^*$. 
We see that the qualitative behavior of $m_s^*$ as a function of $T$ 
(at fixed low $\mu$) is similar to that of
$m_u^*$. Yet this is not the case with the 
$\mu$ dependence of $m_s^*$ at fixed low $T$ where there appear two gaps
(see lower panel). The first gap comes from the effect of $m_u^*$, the
second one comes from that of $m_s^*$.
We noted that the critical chemical potential is less than the
constituent quark mass at $T=0$ in the DR for $2\leq D<3$ at the
chiral limit \cite{Inagaki:1994ec}.

Other three parameter sets, namely Cases $\chi_{170}$, $\chi_{179}$
and $\chi_{179}^{\rm HD}$, lead qualitatively to similar results.
However, it is worth mentioning that the Case $\chi_{179}^{\rm HD}$,
although reproduces proper values for the meson properties discussed
in \cite{Inagaki:2010nb}, shows quite different quantitative behavior.
The constituent quark masses are one order of magnitude larger than
the ones in the other parameter sets. This leads to unphysical
results for the chiral phase transition; the transition temperature
becomes of the order of $1$GeV which is much larger than the
physically expected value. This may indicate that the parameters of
the Case $\chi_{179}^{\rm HD}$ are not suitable to describe the chiral
dynamics in this model. To explicitly compare these parameter sets,
we align the numerical results for the constituent quark mass, $m_u^*$,
in Tab.~\ref{constituent}.
\begin{table} [!h]
 \caption{Constituent quark mass $m_u^*$ [MeV].}
 \label{constituent}
 \begin{ruledtabular}
 \begin{tabular}{lccccc}
 $m_u$ & Case $m_\eta^{\rm LD}$ & Case $m_{\eta}$ & Case $\chi_{170}$ 
       & Case $\chi_{179}$ & Case $\chi_{179}^{\rm HD}$ \\
 \hline %
 $3.0$ & $-467$ & $-597$ & $-453$ & $-455$ & $-1942$ \\
 $4.0$ & $-460$ & $-623$ & $-439$ & $-442$ & $-2260$ \\
 $5.0$ & $-453$ & $-662$ & $-423$ & $-427$ & $-$ \\
 $5.5$ & $-$ & $-679$ & $-415$ & $-420$ & $-$ \\
 $6.0$ & $-$ & $-696$ & $-408$ & $-412$ & $-$ 
 \end{tabular}
 \end{ruledtabular}
\end{table}
There one can clearly see that the values in the Case $\chi_{179}^{\rm HD}$
are considerably larger than the ones in the other parameter sets.
Since the model predictions in the Cases $\chi_{170}$ and
$\chi_{179}$ are almost the same due to the similar values of $m_u^*$
and the model parameters, we shall only show the results in the Case
$\chi_{170}$. 
In what follows, therefore, we will focus on the Cases $m_{\eta}^{\rm LD}$, 
$m_\eta$ and $\chi_{170}$.  

\subsection{$m^*$ in the cutoff regularization scheme}
Here we present the results obtained in the widely used sharp
cutoff scheme  to make comparison with our results of the DR.

\begin{figure}
  \begin{center}
    \includegraphics[width=3.2in,keepaspectratio]{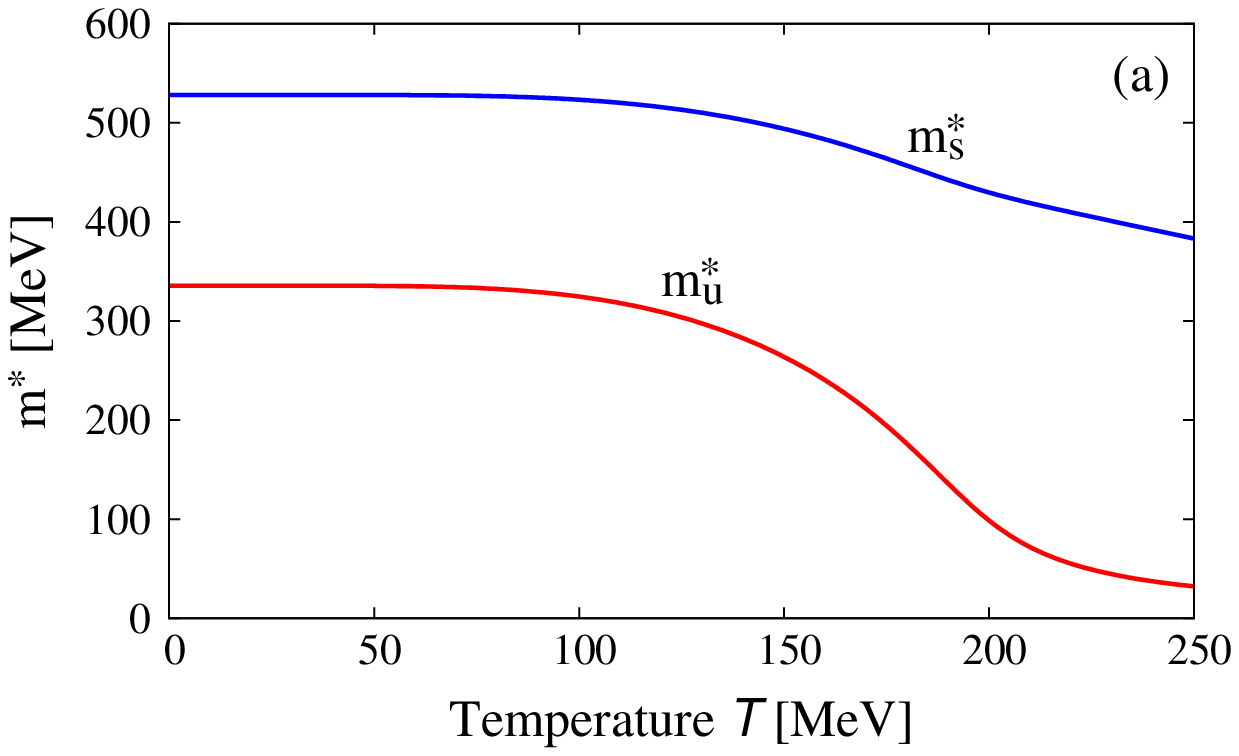}
    \includegraphics[width=3.2in,keepaspectratio]{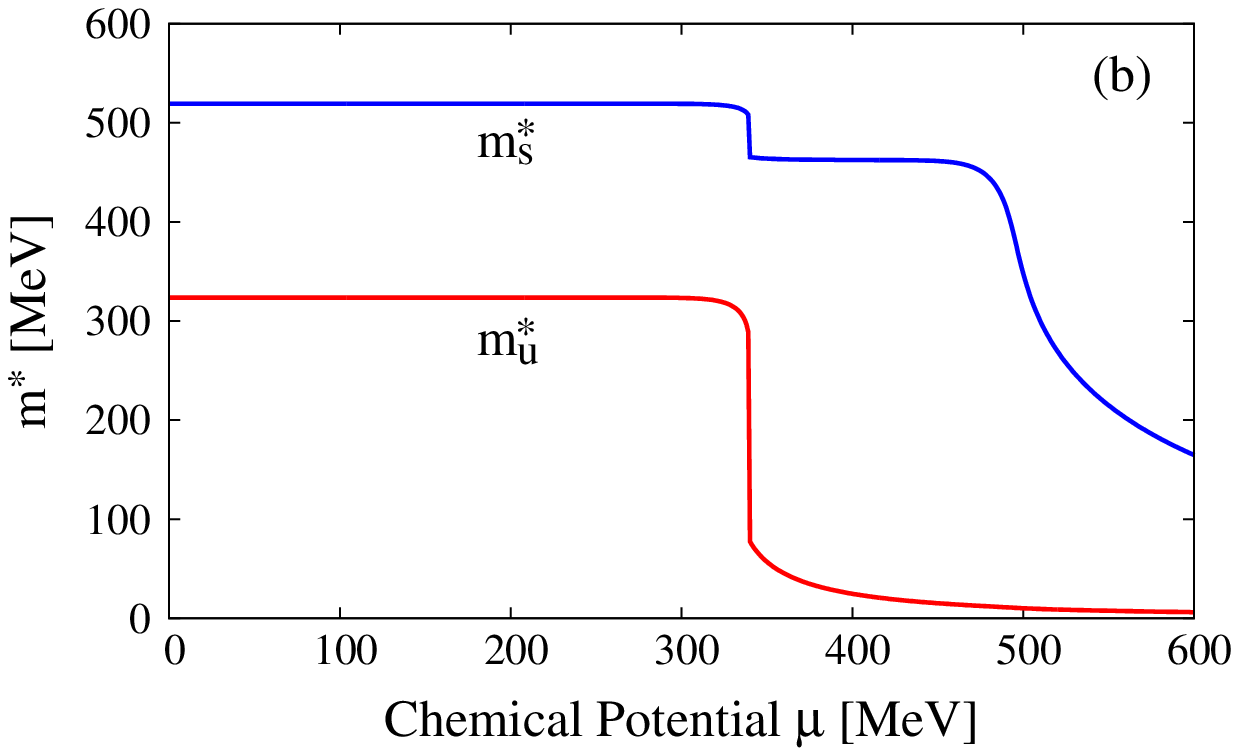}
  \end{center}
  \vspace{-0.3cm}
  \caption{$m_u^*$ and $m_s^*$ in the case of cutoff for $m_u=5.5$MeV.
  (a) finite $T$ and $\mu=0$.
  (b) finite $\mu$ and $T=10$MeV.}
  \label{constituent_cut}
\end{figure}
\begin{figure}
  \begin{center}
    \includegraphics[width=3.2in,keepaspectratio]{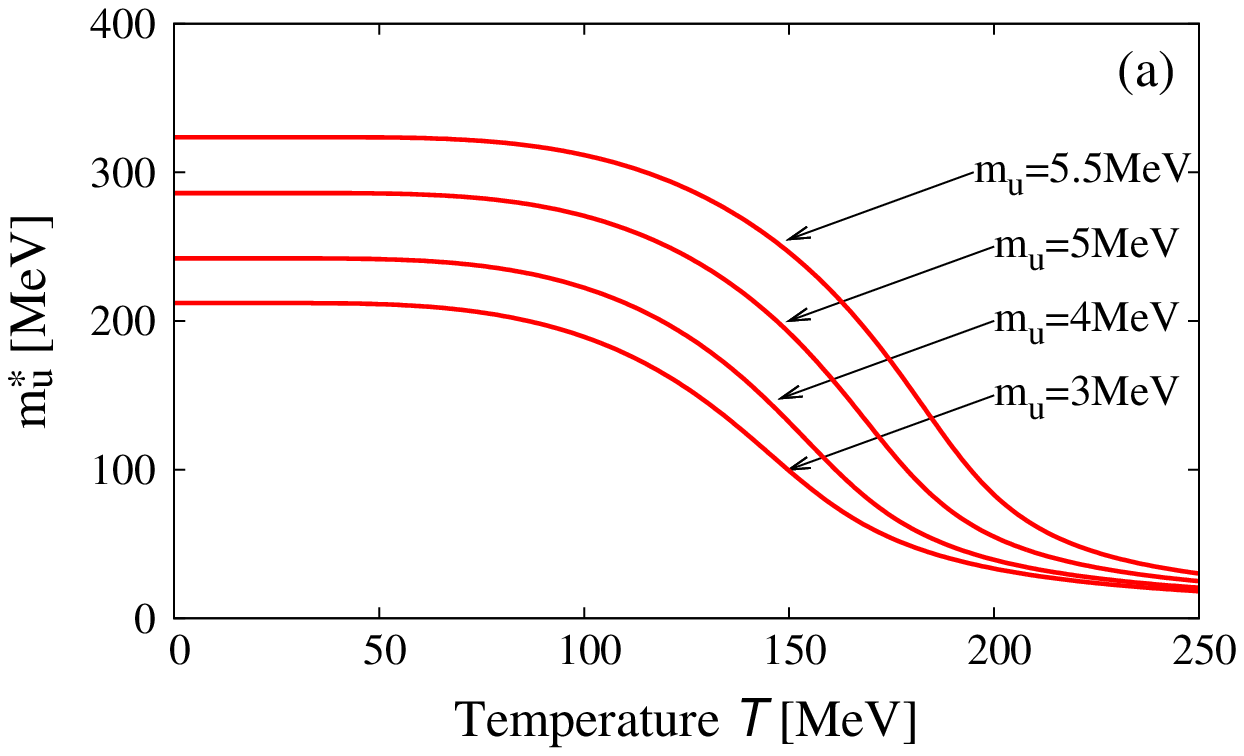}
    \includegraphics[width=3.2in,keepaspectratio]{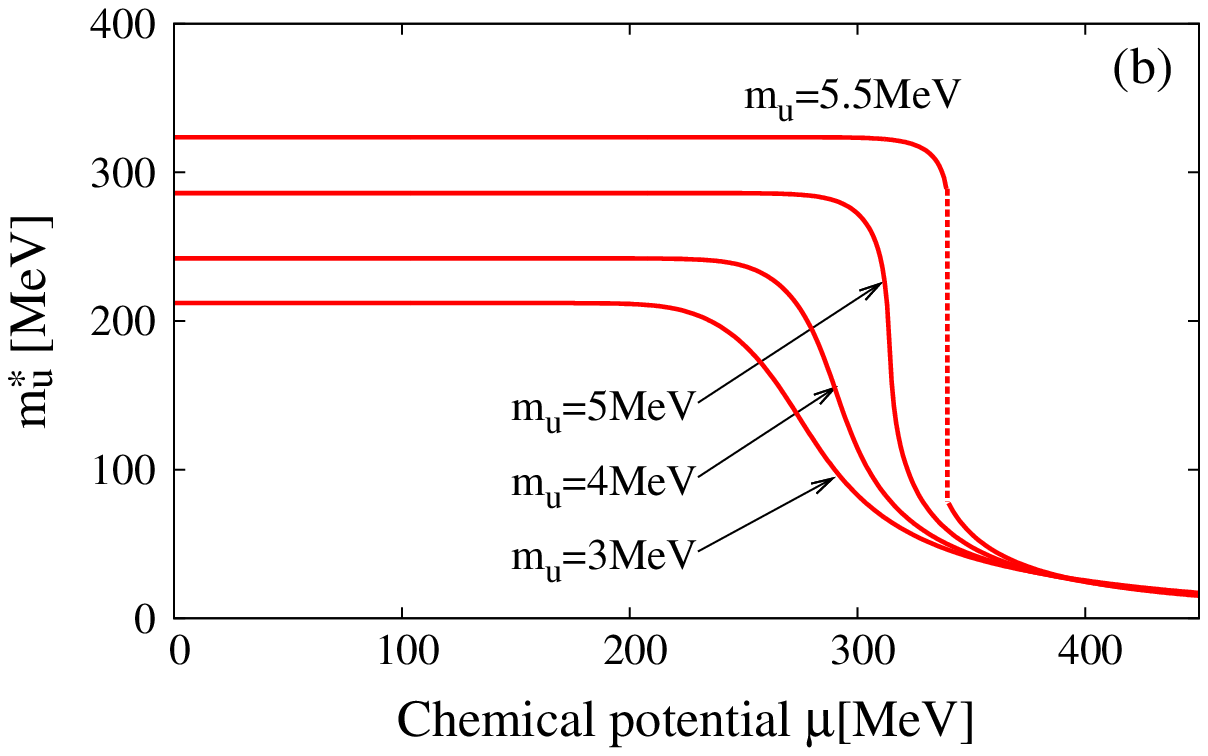}
  \end{center}
  \vspace{-0.3cm}
  \caption{$m_u^*$ in the case of cutoff. (a) finite $T$ and $\mu=0$. 
            (b) finite $\mu$ and $T=10$MeV.}
  \label{Mu_cut}
\end{figure}
\begin{figure}
  \begin{center}
    \includegraphics[width=3.2in,keepaspectratio]{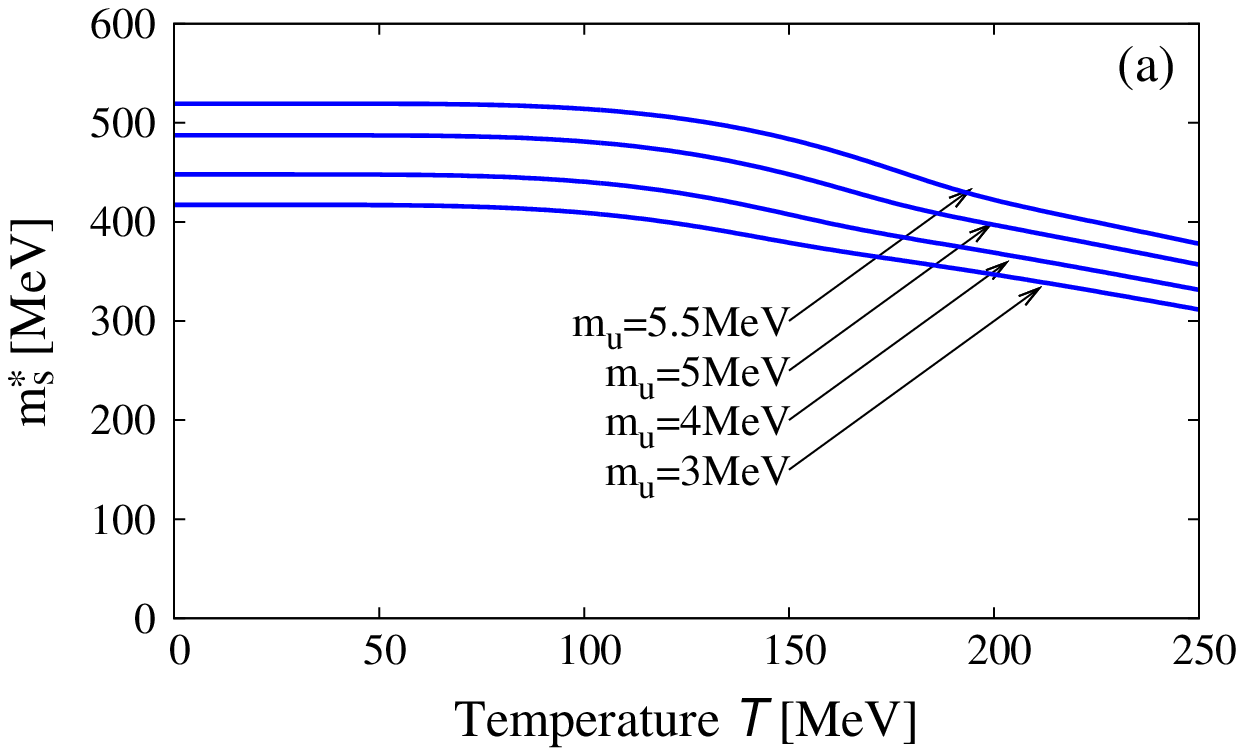}
    \includegraphics[width=3.2in,keepaspectratio]{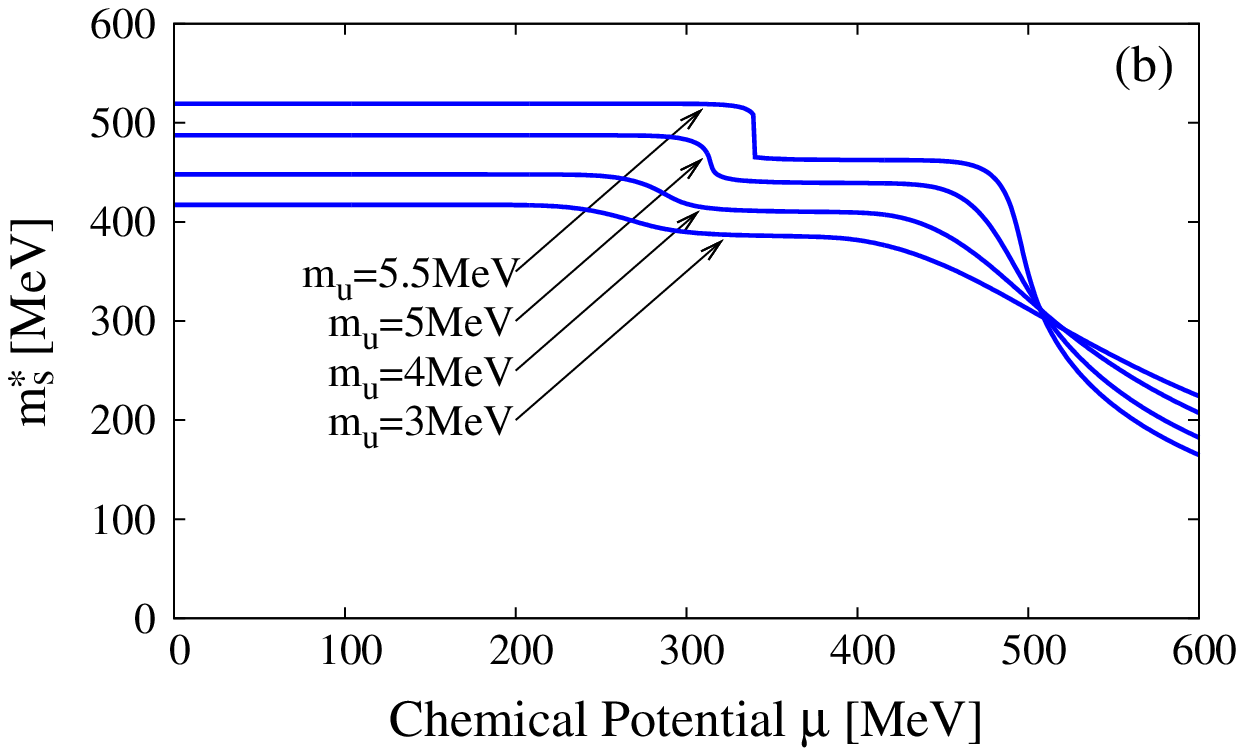}
  \end{center}
  \vspace{-0.3cm}
  \caption{$m_s^*$ in the case of cutoff. (a) finite $T$ and $\mu=0$. 
            (b) finite $\mu$ and $T=10$MeV.}
  \label{Ms_cut}
\end{figure}

The numerical results for the constituent quark masses $m_u^*$ and
$m_s^*$ for the case $m_u=5.5$MeV are shown in Fig.~\ref{constituent_cut}.
It is found that the $T$ and $\mu$ dependences are similar to the DR 
case. Fig.~\ref{Mu_cut} shows qualitative
difference between the cases of $m_u=5.5$MeV, and, $m_u=3$, $4$, $5$MeV.
$\mu$ dependence of the constituent quark mass $m_u^*$ 
has discontinuity in the case of $m_u=5.5$MeV, while $m_u^*$ decreases
continuously in the cases of $m_u=3$, $4$, $5$MeV. Thus, we have the
phase transition of the first order in the case of $m_u=5.5$MeV and
crossover in the other cases. This is a critical qualitative difference
between the dimensional and cutoff regularizations.

Fig.~\ref{Ms_cut} displays the corresponding results for $m_s^*$.
We note that the $T$ dependence of $m_s^*$ resembles each other in the
two regularizations, which is seen in Figs.~\ref{Ms_eta}(a) and
\ref{Ms_cut}(a). However, as seen from Figs.~\ref{Ms_eta}(b) and
\ref{Ms_cut}(b), the $\mu$ dependence of $m_s^*$ is different. 
The two gaps in the cutoff regularization
are smaller and smoother than those in the dimensional scheme.

\section{Meson properties}
\label{meson}
In this section, we will show the meson properties, \{$m_{\pi}$, $f_{\pi}$,
$m_{\mathrm K}$, $m_\eta$, $m_{\eta^{\prime}}$\}, in the two regularizations.
We present the Cases $m_{\eta}$ and $\chi_{170}$ with $m_u=5.5$MeV,
and $m_{\eta}^{\rm LD}$ with $m_u=3$MeV. The reason of these choices is
that the case with $m_u=5.5$MeV is frequently used value in the cutoff
regularization~\cite{Hatsuda:1994pi, Rehberg:1995kh}. As the Case
$m_{\eta}^{\rm LD}$ does not have solutions at $m_u=5.5$MeV, so we select
the $m_u=3$MeV case in which the critical temperature of the chiral
phase transition is close to empirical data (see,
Sec.~\ref{critical}).

\subsection{Meson properties in DR} 
\label{sec_meson_dim}
Here the numerical results of the meson properties are presented for the
Cases $\chi_{170}$, $m_{\eta}$ and $m_{\eta}^{\rm LD}$ with some discussion
on their behavior.

\begin{figure}
  \begin{center}
    \includegraphics[width=3.2in,keepaspectratio]{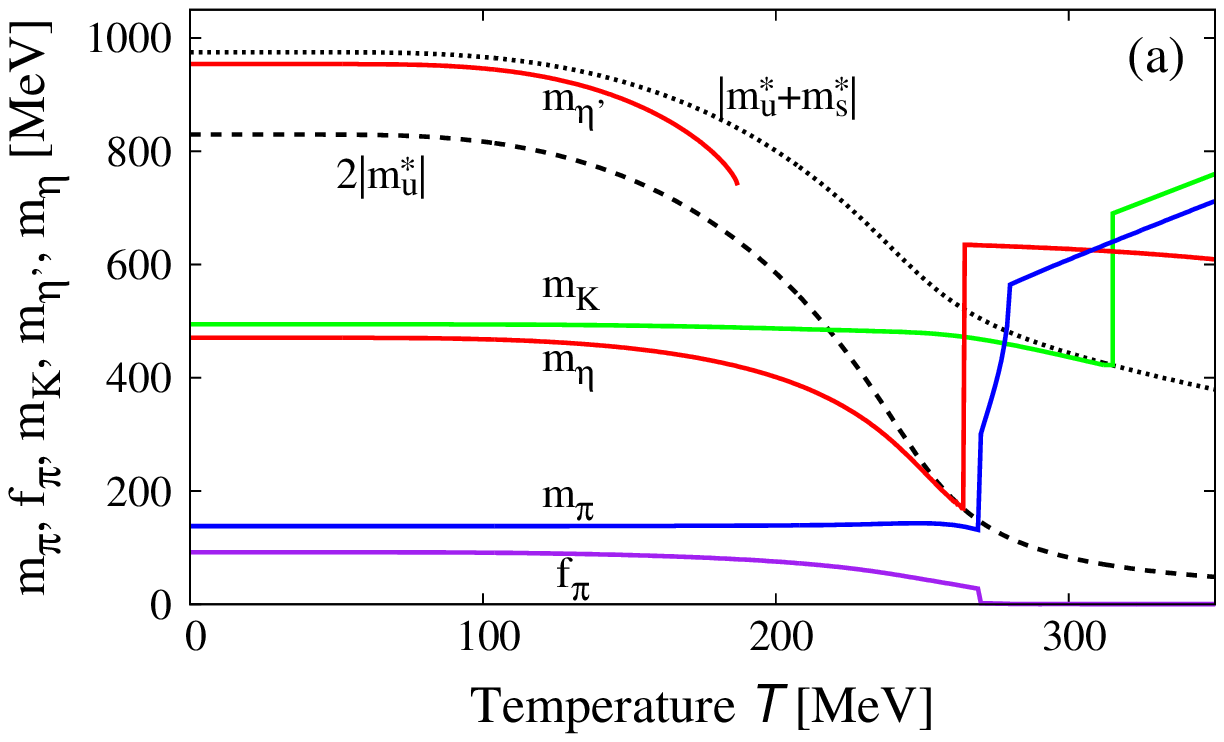}
    \includegraphics[width=3.2in,keepaspectratio]{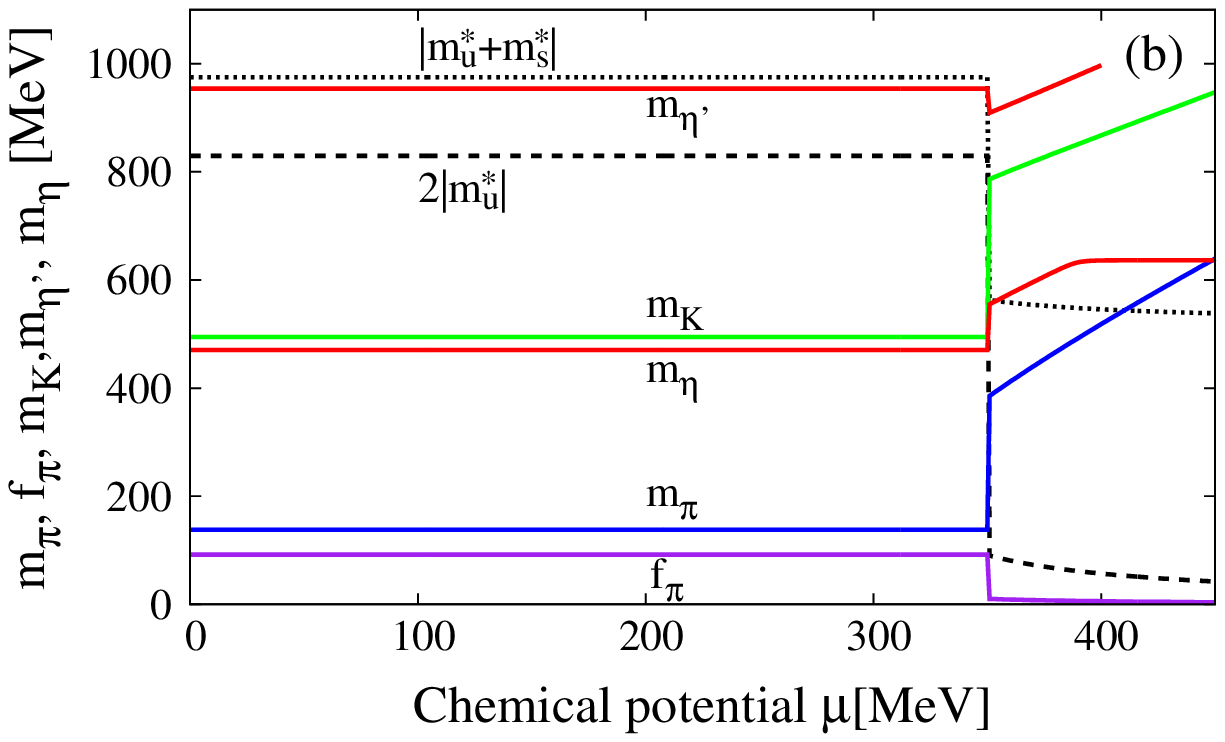}
  \end{center}
  \vspace{-0.3cm}
  \caption{Case $\chi_{170}$ with $m_u=5.5$MeV. (a) finite $T$ and $\mu=0$.
           (b) finite $\mu$ and $T=10$MeV.}
  \label{fit_chi17055}
\end{figure}
Fig.~\ref{fit_chi17055} shows the results of the meson properties for (a)
finite $T$ and $\mu=0$, and (b) finite $\mu$ and $T=10$MeV in the Case
$\chi_{170}$ with $m_u=5.5$MeV. As one can observe from the upper panel,
$m_{\pi}$ and $m_{\eta}$ stay almost constant until they cross $2|m_u^*|$,
then suddenly jump. In the similar manner, $m_{\mathrm K}$ is nearly
constant below the value $|m_u^* + m_s^*|$, and it enhances when $m_{\mathrm K}$
becomes $|m_u^* + m_s^*|$. $f_{\pi}$ decreases monotonously and becomes
negligibly small after $m_\pi$ exceeds $2|m_u^*|$. At low $T$ the
dependence of meson properties on $\mu$ has sharp discontinuities as
seen in Fig.~\ref{fit_chi17055}(b), where the masses $m_{\pi}$,
$m_{\mathrm K}$, $m_{\eta}$ suddenly get larger and $m_{\eta^{\prime}}$,
$f_{\pi}$ falls off. This sharp discontinuities come from the
discontinuous change of the chiral condensate, 
which corresponds to the strong first order phase transition. 
On the other hand in the upper panel the meson masses also show jumps 
as functions of $T$, although the chiral condensates change continuously 
with $T$. This comes from the singular behavior which appears particularly
in the lower dimensions $D<4$. Below, we will discuss
the origin of these discontinuities through the analysis of the solution
for the pion state. 

\begin{figure}
  \begin{center}
    \includegraphics[width=3.2in,keepaspectratio]{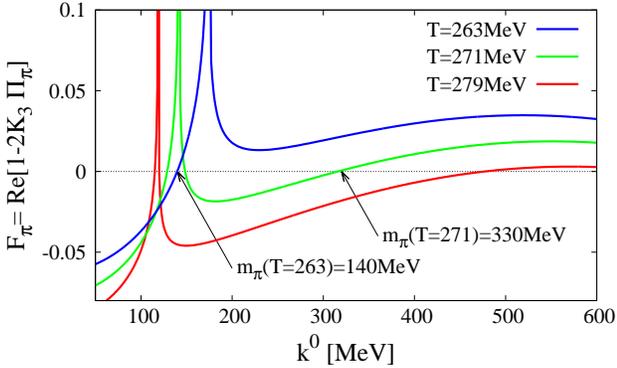}
  \end{center}
  \vspace{-0.3cm}
  \caption{${\mathcal F}_{\pi}$ (denominator of $\Delta_\pi$) 
   for three fixed values of $T$, 263, 271, 279MeV, in the Case 
   $\chi_{170}$ with $m_u=5.5$MeV at $\mu=0$.}
  \label{m_pi_pole}
\end{figure}
In Fig.~\ref{m_pi_pole} the function 
${\mathcal F}_{\pi}(k^0)={\rm Re}[1-2K_3\Pi_\pi(k^0)]$ is plotted
which determines the pion mass through the condition
${\mathcal F}_{\pi}(k^0)=0$ (see App.~\ref{app_pk_mass}).
Note that all the lines diverge at $k^0=2|m_u^*|$. 
Due to this divergent behavior of ${\mathcal F}_{\pi}(k^0)$, 
there is a jump in the dependence of the pion mass on $T$.
Namely, we find the unique
solution around $k^0 \simeq 140$MeV at $T=263$MeV. However, there appear
three candidates for the solution at $T=271$MeV; two of them are located
near the divergent point and the third is larger.
We select the larger value, $k^0 \simeq 330$MeV, because the pion
should acquire a larger mass at high $T$ where the chiral symmetry is expected
to be restored. This is the reason why the discontinuous change of
$m_{\mathrm \pi}$ at some $T$, is not originated from the
sudden change of the chiral condensate $\phi_u$. In App.~\ref{app_pk_mass},
we give the technical details on the divergence in ${\mathcal F}_{\pi}$.

\begin{figure}[!h]
  \begin{center}
    \includegraphics[width=3.2in,keepaspectratio]{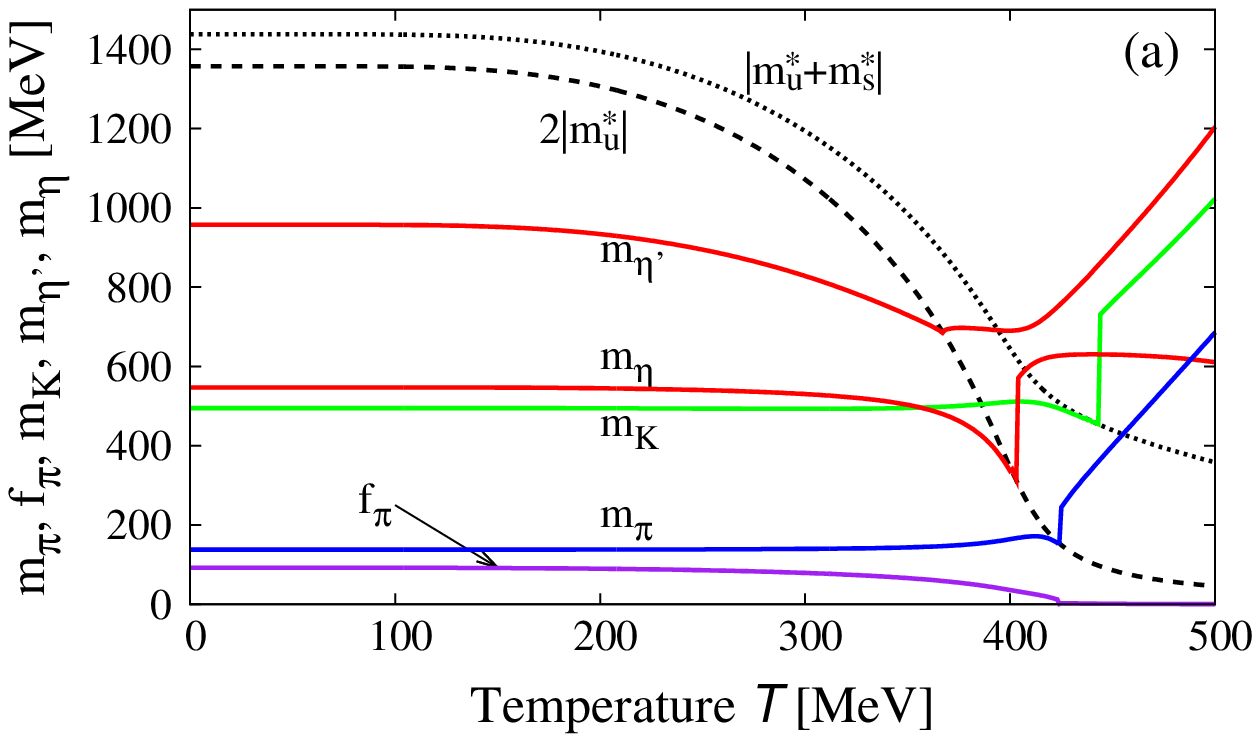}
    \includegraphics[width=3.2in,keepaspectratio]{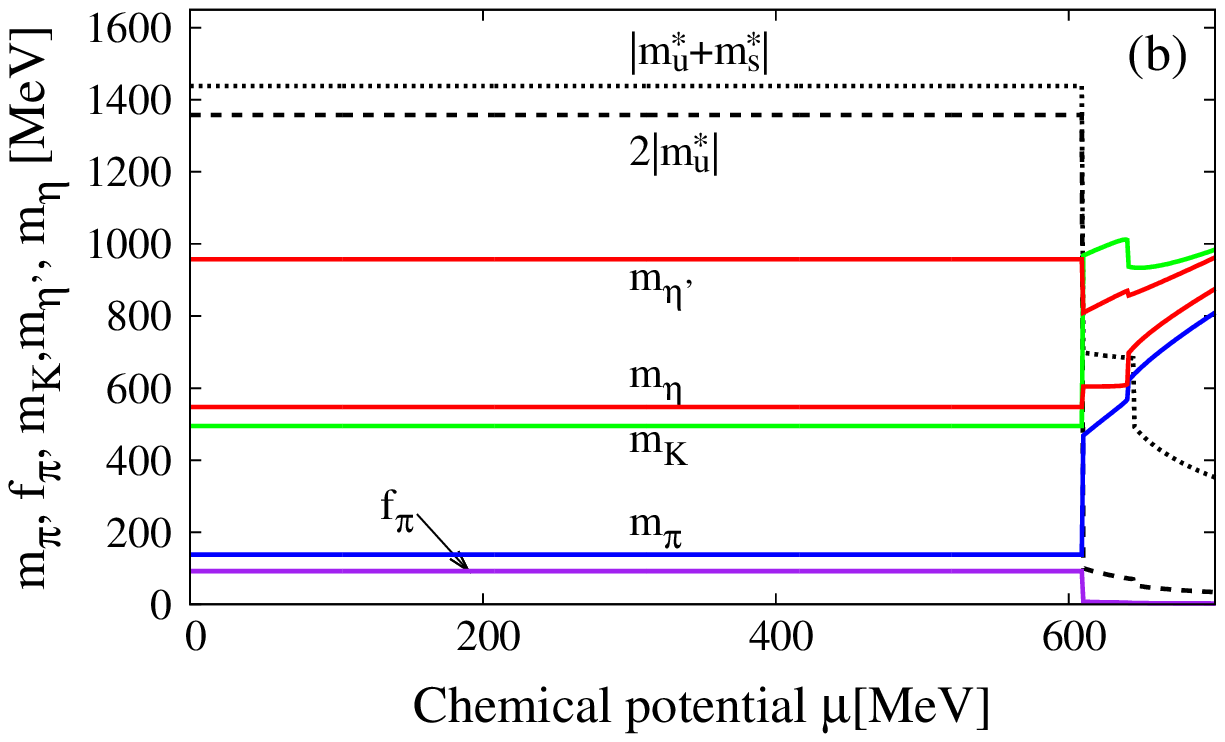}
  \end{center}
  \vspace{-0.3cm}
  \caption{Case $m_{\eta}$ with $m_u=5.5$MeV. (a) finite $T$ and $\mu=0$.
           (b) finite $\mu$ and $T=10$MeV.}
  \label{meson_eta55}
\end{figure}
\begin{figure}[!h]
  \begin{center}
    \includegraphics[width=3.2in,keepaspectratio]{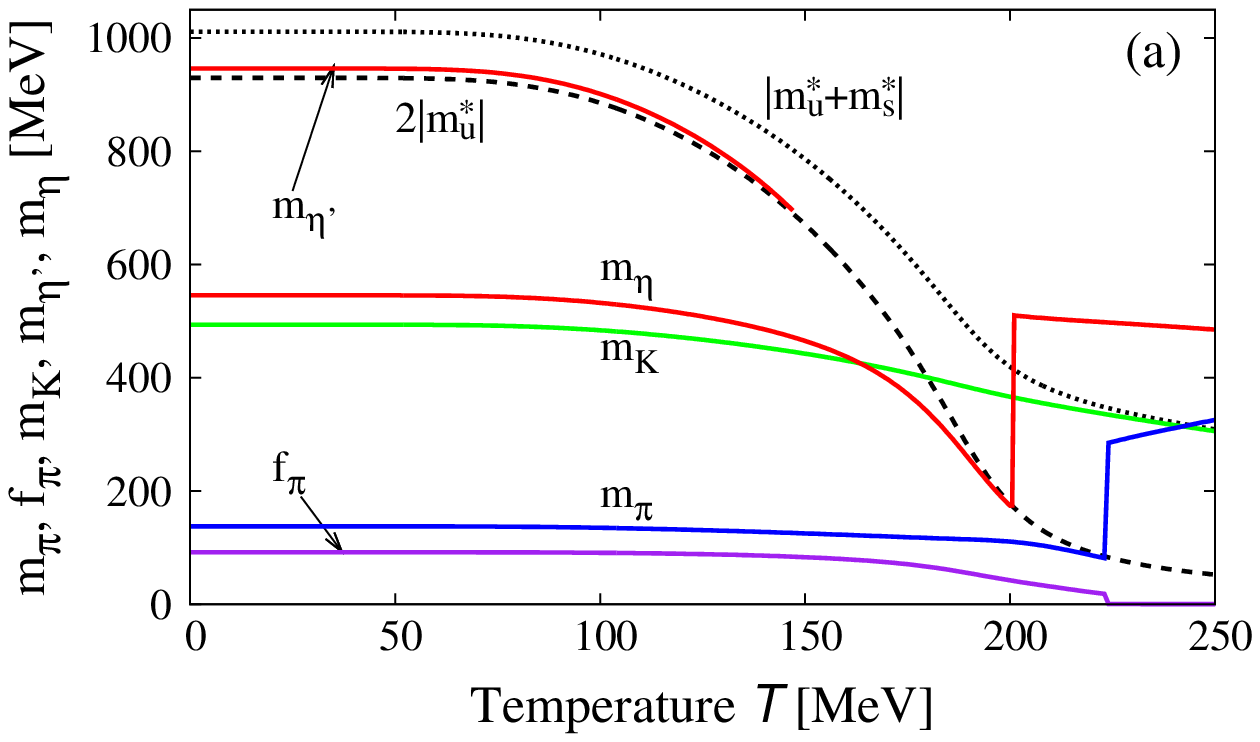}
    \includegraphics[width=3.2in,keepaspectratio]{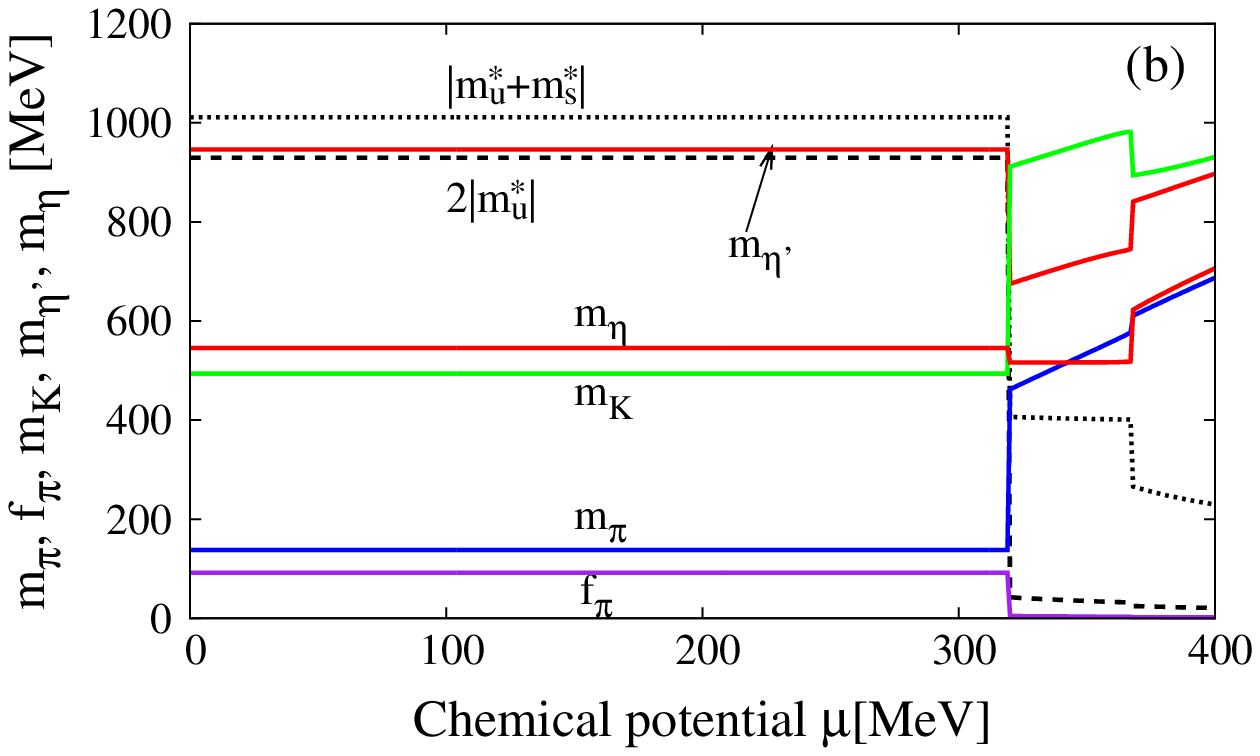}
  \end{center}
  \vspace{-0.5cm}
  \caption{Case $m_{\eta}^{\rm LD}$ with $m_u=3$MeV. (a) finite $T$ and $\mu=0$.
           (b) finite $\mu$ and $T=10$MeV.}
  \label{meson_eta30}
\end{figure}

In Figs.~\ref{meson_eta55} and \ref{meson_eta30}, we show the results for
the Case $m_\eta$ and $m_\eta^{\rm LD}$. We observe the similar pictures in
these two cases; smooth dependence on $T$ at zero $\mu$, and similar
discontinuities in the dependence on $\mu$ at low $T$.  As is seen in
Figs.~\ref{meson_eta55}(b) and \ref{meson_eta30}(b), there are two
discontinuities. The first discontinuity comes from the effect of the gap
for $m_u^*$, and the second discontinuity comes from the one for $m_s^*$.
These gaps are confirmed in Figs.~\ref{Mu_eta}(b) and \ref{Ms_eta}(b). It
is interesting to note that $m_{\eta'}$ survives at high temperature region
in Fig.~\ref{meson_eta55}(a). This behavior may be attributed to the fact
that $m_{\eta'}$ is smaller than the threshold $2|m_u^*|$ at $T=0$.      

We have also calculated meson properties for the cases of different
$m_u$, and found that the differences between various cases are rather
nominal, therefore we showed the results for the case of 
a particular $m_u$.

\subsection{Meson properties in cutoff regularization}  
 Here we present the results for the meson properties
in the cutoff case.

\begin{figure}
  \begin{center}
    \includegraphics[width=3.2in,keepaspectratio]{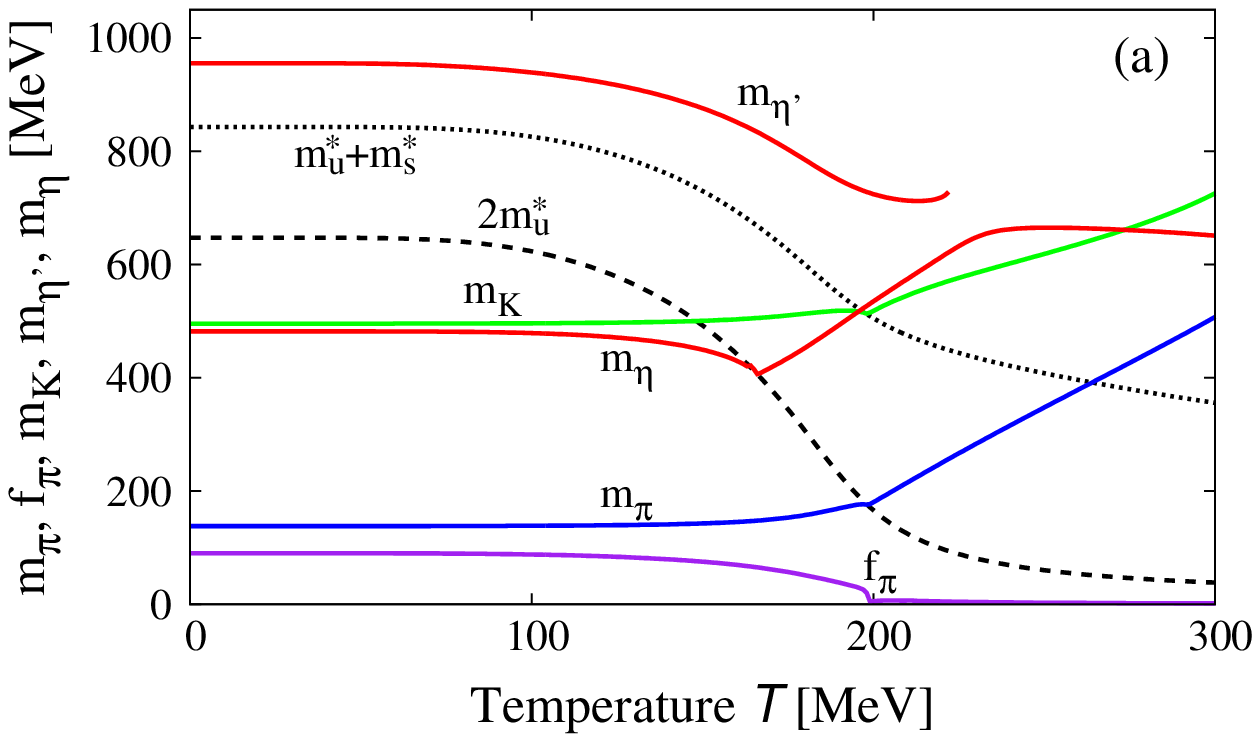}
    \includegraphics[width=3.2in,keepaspectratio]{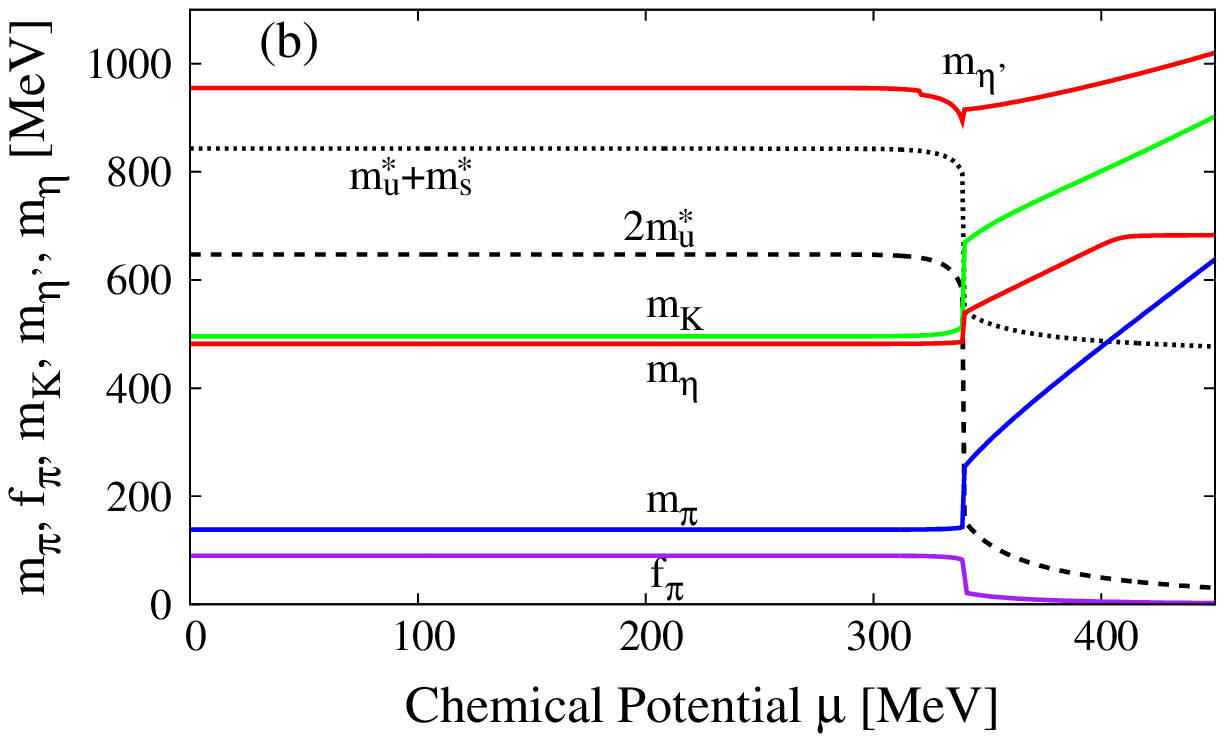}
  \end{center}
  \vspace{-0.3cm}
  \caption{Meson properties in the case of cutoff for $m_u=5.5$MeV.
  (a) finite $T$ and $\mu=0$.
  (b) finite $\mu$ and $T=10$MeV.}
  \label{meson_cut}
\end{figure}
Fig.~\ref{meson_cut} shows the dependence of the meson properties on
$T$ and $\mu$. The results in the two regularizations 
look similar to each other. The most noticeable difference
is in that the curves are rather smooth in the cutoff case. There are no
discontinuities for low $\mu$ in Fig.~\ref{meson_cut}(a), because
Eq.~(\ref{I_ij0}) does not include divergence for $D=4$, and the solution
on the real axis survives at high temperature region. In Fig.~%
\ref{meson_cut}(b) the gap at $\mu\simeq 340$MeV is smaller than the
gap of the DR  case. This can be understood from
the fact that the gap of $m_u^*$ in the cutoff scheme is smaller than
the one in the DR scheme (see, Figs.~%
\ref{Mu_eta}(b), \ref{Mu_cut}(b)).

\section{Topological susceptibility}
\label{topo}

The topological susceptibility $\chi$ is an important quantity because
it is intimately related to the chiral and \ua symmetry of QCD. We show the
results of $\chi$ in the dimensional and cutoff regularizations in this
section. 

\begin{figure}
  \begin{center}
    \includegraphics[width=3.2in,keepaspectratio]{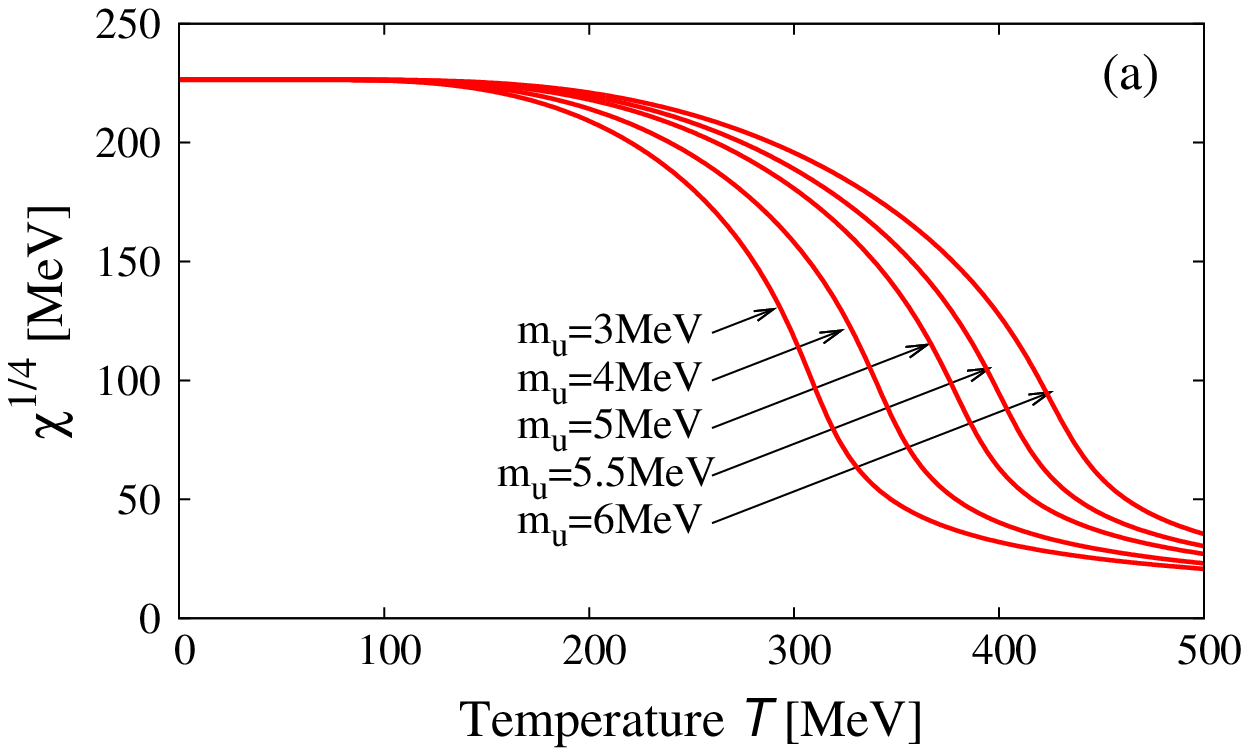}
    \includegraphics[width=3.2in,keepaspectratio]{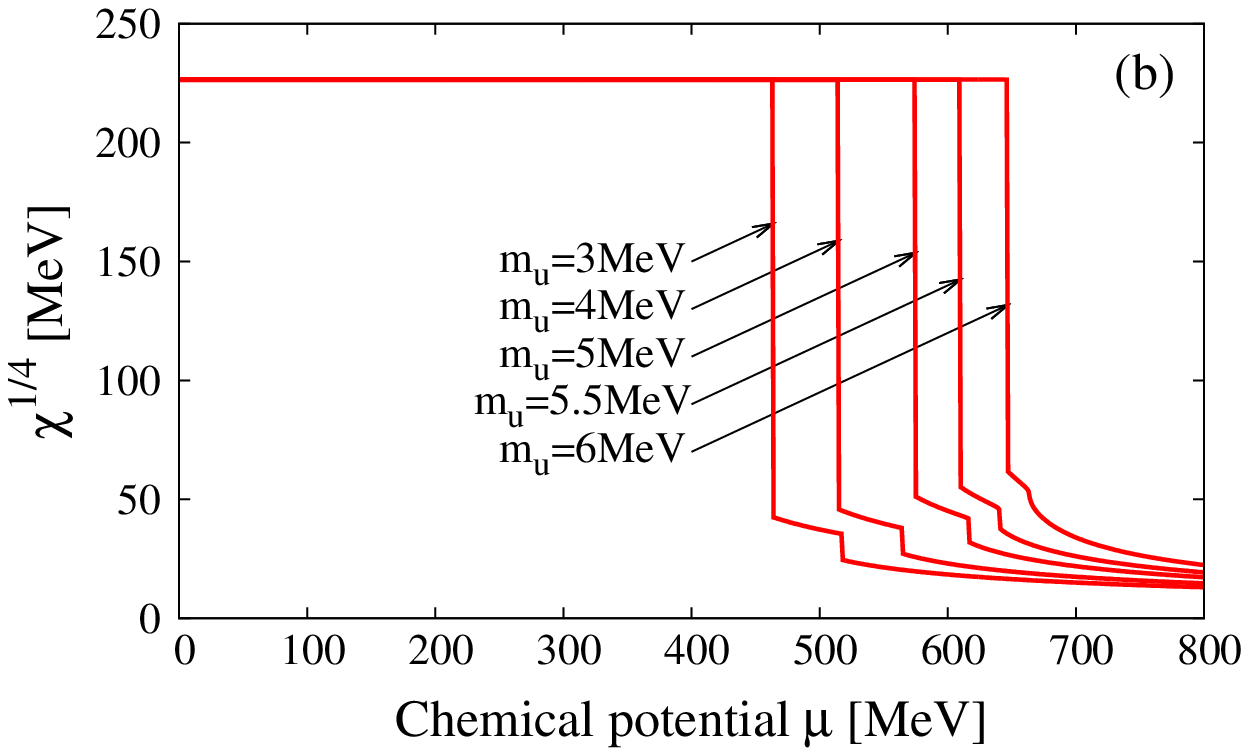}
  \end{center}
  \vspace{-0.3cm}
  \caption{Case $m_{\eta}$: Topological susceptibility $\chi^{1/4}$.
           (a) finite $T$ and $\mu=0$.
           (b) finite $\mu$ and $T=10$MeV.}
\label{topo_eta}
\end{figure}
\begin{figure}
  \begin{center}
    \includegraphics[width=3.2in,keepaspectratio]{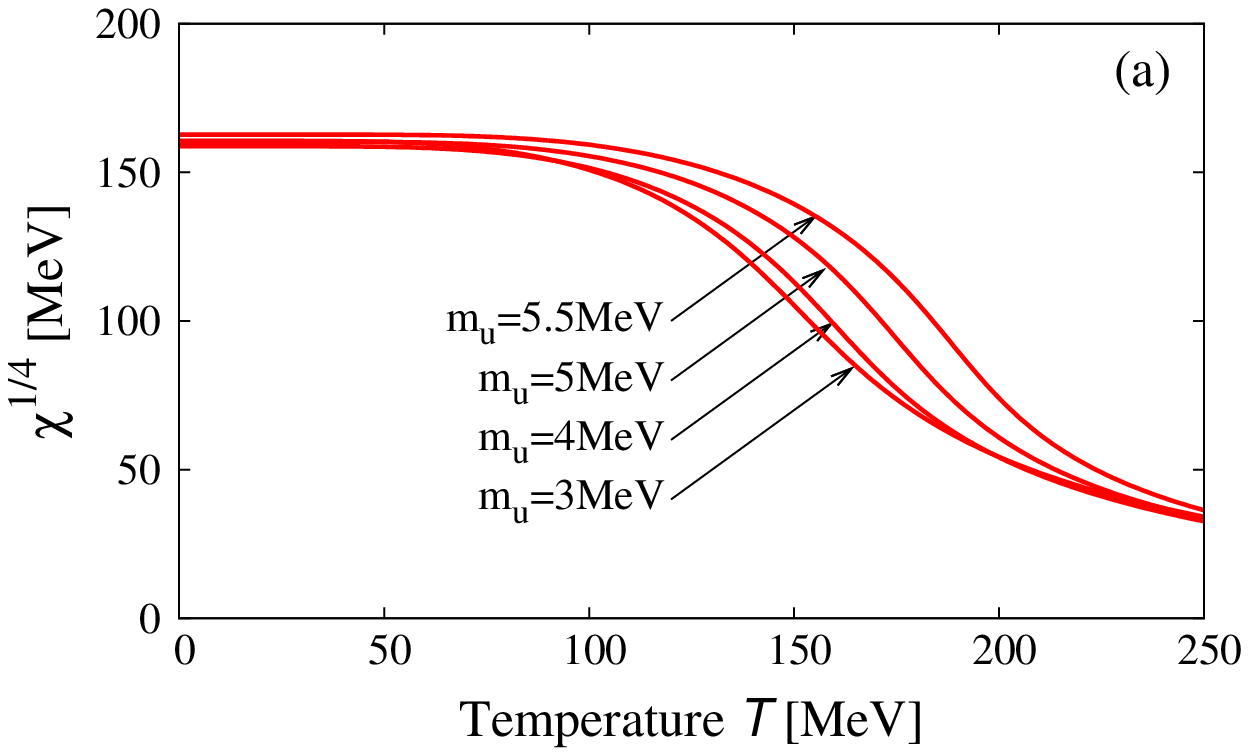}
    \includegraphics[width=3.2in,keepaspectratio]{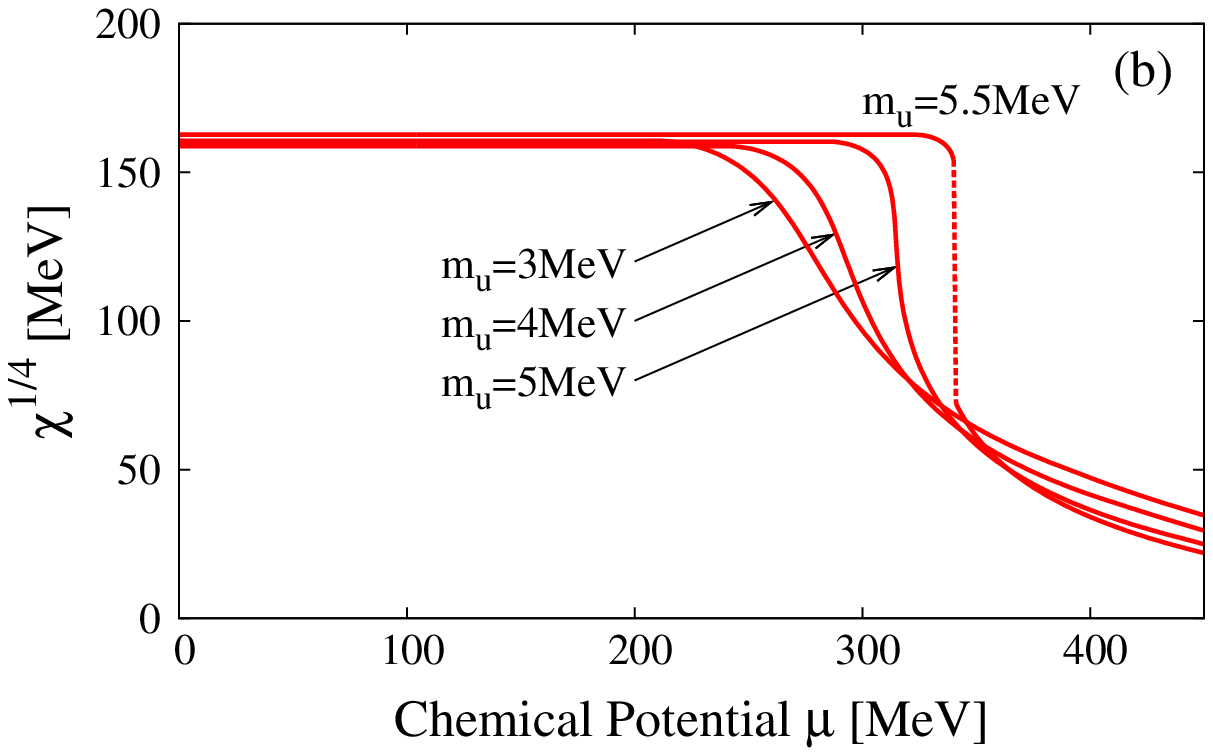}
  \end{center}
  \vspace{-0.3cm}
  \caption{Topological susceptibility in the case of cutoff.
           (a) finite $T$ and $\mu=0$.
           (b) finite $\mu$ and $T=10$MeV.}
\label{topo_cut}
\end{figure}
Figs.~\ref{topo_eta} and \ref{topo_cut} display the curves for
the topological susceptibility $\chi$ as the functions of $T$ or $\mu$
which resemble the curves for $m_u^*$
in Figs.~\ref{Mu_eta} and \ref{Mu_cut} respectively. Thus we confirm that the
topological susceptibility is influenced indeed by the
chiral symmetry breaking.

\section{Critical temperature}
\label{critical}
We have seen that the chiral condensate and the topological
susceptibility always show discontinuous changes at low temperature
in the DR, which indicates the first order
phase transition. The tendency remains in all cases
with different parameter sets. On the other hand, the order of
transition is different for the different parameter sets in the cutoff
case as confirmed in Fig.~\ref{Mu_cut}(b) and \ref{topo_cut}(b).
This may indicate that the transition pattern crucially depends on
the regularization procedures. Therefore it is interesting to study
the values of the critical temperature $T_c$ for various cases both
in dimensional and cutoff regularizations. Here we focus on $T_c$ at
$\mu=0$, and make the comparison among the two regularization and the
lattice QCD simulation.

There are several ways to define the critical temperatures. In this
paper, we shall employ two definitions given by the maxima of
\begin{align}
  \frac{\partial \phi_u}{\partial T},
  \quad {\rm and} \quad
  -\frac{\partial \chi^{1/4}}{\partial T}.
\label{tc_def}
\end{align}
The resulting critical temperatures will be denoted by,
$T_{c}^{(\phi)}$ and $T_{c}^{(\chi)}$, respectively.

\begin{table}
  \caption{Critical temperature $T_c^{(\phi)}$.}
  \label{T_phi}
  \begin{ruledtabular}
  \begin{tabular}{lcccc}
  $m_u$  & $m_\eta^{\rm LD}$ & $m_{\eta}$  & $\chi_{170}$  & Cutoff  \\
  \hline %
  $3.0$  & $184$ &  $304$  & $253$ & $146$ \\
  $4.0$  & none  & $336$  & $249$ & $155$ \\
  $5.0$  & none  & $374$  & $243$ & $170$ \\
  $5.5$  & $-$   & $395$  & $240$ & $184$ \\
  $6.0$  & $-$   & $419$  & $237$  & $-$
  \end{tabular}
  \end{ruledtabular}
\end{table}
%
\begin{table}
  \caption{Critical temperature $T_c^{(\chi)}$.}
  \label{T_chi}
  \begin{ruledtabular}
  \begin{tabular}{lcccc}
  $m_u$  & $m_{\eta}^{\rm LD}$ & $m_\eta$  & $\chi_{170}$ & Cutoff  \\
  \hline %
  $3.0$ & 188  & $309$  & $259$ & $152$ \\
  $4.0$ & none & $341$  & $255$ & $161$ \\
  $5.0$ & none & $379$  & $250$ & $175$ \\
  $5.5$ & $-$  & $401$  & $246$ & $189$ \\
  $6.0$ & $-$  & $425$  & $244$ &  $-$
  \end{tabular}
  \end{ruledtabular}
\end{table}
In Tab.~\ref{T_phi} the numerical results for the critical
temperature are aligned for various parameter sets. In the Case 
$m_{\eta}^{\rm LD}$ with $m_u=4$ and $5$MeV, we can not find 
a physically meaningful behavior
for the phase transition; the absolute value of $m_u^*$ becomes larger
with increasing $T$. That is why these cases are marked by 
``none'' in the table.

Note that the values of $T_c$ in the DR except 
for the Case $m_\eta^{\rm LD}$ are larger than ones in the cutoff case. 
The Case $m_\eta^{\rm LD}$ at $m_u=3$MeV leads almost to the same value of 
$T_c$ as in the cutoff case at $m_u=5.5$MeV which is comparable with the 
result in the lattice QCD simulation, $150-200$MeV \cite{Fukushima:2010bq}. 
We see that $T_c$ becomes larger
with increasing $m_u$ in the Case $m_{\eta}$ while $T_c$ does not change
drastically with $m_u$ in the Case $\chi_{170}$. This can be 
explained by the values of $m_u^*$.

\section{Summary and discussions}
\label{conclusion}
In this paper we studied the nonet meson characteristics, $m_{\pi}$,
$f_{\pi}$, $m_{\mathrm K}$, $m_{\eta}$, $m_{\eta^{\prime}}$, and $\chi$ at
finite temperature and chemical potential in the NJL model with
DR.   Many of meson properties show reasonable
behavior, which is similar to one
obtained in the model with the cutoff regularization.

We examined the behavior of the constituent quark
masses, $m_u^*$ and $m_s^*$, in the two regularizations. $T$ dependence
of the constituent quark masses at low $\mu$ 
does not depend essentially on the way of regularization. 
However, $\mu$ dependence of the constituent quark masses at low $T$ 
depends on the regularization way around the critical 
chemical potential. In the DR, the phase transition 
at low $T$ is always of the first order for various choices of $m_u$.
This tendency is consistent with the current consensus about
the transition to be of the first order in the chiral limit~%
\cite{Pisarski:1983ms}.
On the other
hand in the cutoff case, the gap at the critical chemical potential
becomes smaller with decreasing value of $m_u$, and it eventually vanishes
resulting in crossover transition (see, Fig.~\ref{Mu_cut}). This happens due
to the fact that the larger cutoff $\Lambda$ diminishes the coupling
strength which leads to the smoother change of the order parameters.
Decreasing the coupling strength weakens nonperturbative effects. This
explains the effect of the UV-cutoff which is also seen in the 
work~\cite{Chen:2009mv} as the effect of the cutoff in the temporal direction.

Using the obtained values of the constituent quark masses, we evaluated
meson properties at finite $T$ and/or $\mu$ in the two 
regularizations. $T$ dependence of the meson masses at low $\mu$ are 
similar to each other except around $m_{\mathrm P} \simeq |m_i^*+m_j^*|$. 
Since the singularity of the pion self-energy $\Pi_{\mathrm P}(k^2)$ 
appears at $k^2=(m_i^*+m_j^*)^2$ for $D<4$, the behavior of meson masses 
becomes discontinuous in the DR. $\mu$ dependence 
of the meson masses at low $T$ also does not essentially depend on the way 
of regularization except at the region of the critical chemical potential. 
Since the gap of $m_u^*$ in the DR case is larger 
than one in the cutoff case, the gap of the meson mass becomes large. 
The behavior of $f_\pi$ as the functions of $T$ or $\mu$ in the DR 
case resembles one in the cutoff case. We also evaluated
the topological susceptibility $\chi$ as the functions of $T$ or $\mu$ in 
the two regularizations. The behavior of $\chi$ is similar to one 
of $m_u^*$.
 
In the DR, we examined the three parameter set cases:
$\chi_{170},\ m_\eta$ and $m_\eta^{\rm LD}$.
Parameters of each set are fixed by fitting physical quantities at 
$T, \mu=0$ \cite{Inagaki:2010nb}. 
Most of the properties concerning meson masses and decay constants are 
similar to each other in the three cases. However, the large 
difference is seen in the values of the critical temperature and of the 
critical chemical potential. One of the sources of this difference may be 
the strength of couplings, i.e., increasing the values of the coupling 
strength decreases the values of the critical temperature and the critical 
chemical potential. We actually calculated the critical temperatures, 
$T_c^{(\phi)}$ and $T_c^{(\chi)}$. In the Case $m_\eta^{\rm LD}$ at 
$m_u=3$MeV, the critical temperature is almost the same as in the cutoff 
case at $m_u=5.5$MeV which is comparable to the result in the lattice QCD 
simulation.

The lessons we learned in this paper are twofold: (i) similarities in
the predicted meson properties and (ii) remarkable differences seen in
the order of the chiral phase transition at finite chemical potential.
Regarding the point (i), we have been convinced that the NJL model
predictions do not depend drastically on whether one uses the dimensional
or cutoff regularization procedure.
This may be the regularization independent aspect of the NJL model,
which is intriguing problem to be studied in more detail in future.
As to the point (ii), the results of the paper suggest 
that the critical point in the phase diagram on $T-\mu$ plane
shows different behavior. We think that it is interesting to
study the behavior of the chiral phase transition and the phase diagrams
of the model in more detail, especially in context of 
how the location of the critical point depends on the parameters in the
different regularization schemes.

\begin{acknowledgments}
HK is supported by the grant NSC-99-2811-M-033-017 from National Science
Council (NSC) of Taiwan. 
\end{acknowledgments}

\appendix
\section{Chiral condensate}
\label{app_chiral}
By taking the trace with respect to the Dirac spinor indices in Eq.
(\ref{trace}), we get the chiral condensate
\begin{equation}
  \phi_i = {\rm tr1} \cdot m^*_i \!\! \int \!\!
           \frac{d^{D-1}p}{(2\pi)^{D-1}}\, T\!\!\!
           \sum_{n=-\infty}^{\infty} \frac{-1}{\omega_n^2 + E_i^2}\,,
\end{equation}
where ${\rm tr}1=N_c\cdot2^{D/2}$. After performing the frequency
summation, we arrive at the expression
\begin{align}
  \phi_i = {\rm tr1}\cdot  m^*_i \!\! \int \!\!
           \frac{d^{D-1}p}{(2\pi)^{D-1}} \frac{-1}{2E_i}
           \left[ 1 - \sum_{\pm} f(E^{\pm}_i) \right],
\end{align}
with the Fermi-Dirac distribution $f(E)$ given by
\begin{align}
f(E^{\pm}_i) \equiv \frac{1}{1+e^{\beta E_i^{\pm}}}.
\label{fermi-dirac}
\end{align}

\section{Meson properties}
\label{app_meson}
Here we evaluate meson masses, decay constants and topological
susceptibility. More detailed, self-contained calculations are presented
in the review paper~\cite{Klevansky:1992qe}.

\subsection{$\pi$ and ${\mathrm K}$ masses}
\label{app_pk_mass}
Meson masses are obtained through examining the pole structure of their
propagators. Using the random-phase approximation and the $1/N_c$
expansion, we get the meson propagators of the form~%
\cite{Klevansky:1992qe, Hatsuda:1994pi}  
\begin{equation}
\Delta_{\mathrm P}(k^2) = \frac{2K_\alpha}{1-2K_\alpha\Pi_{\mathrm P}(k^2)} 
+\mbox{O}({N_c}^{-1}),
\label{pro_P}
\end{equation}
where the index ${\mathrm P}(=\pi,\,{\mathrm K})$ denotes the meson
species and $\alpha$ labels the isospin channel. The pole position is
determined by the equation
\begin{equation}
1-2K_\alpha\Pi_{\mathrm P}(x^2)=0
\label{pk_mass}
\end{equation}
where $x^2$ is regarded as the square of a meson mass, i.e., $m_{\pi}^2$
or $m_{\mathrm K}^2$. 

The effective couplings, $K_\alpha$, of mesons become
\begin{align}
  K_3 &\equiv G - \frac{1}{2} K \phi_s, \quad {\rm for}\,\,\,\, \pi^0,
  \label{K_3}\\
  K_6 &\equiv G - \frac{1}{2} K \phi_u, \quad {\rm for}\,\,\,\, 
       {\mathrm K}^0, \bar{\mathrm K}^0.
  \label{K_6}
\end{align}
The self-energy, $\Pi_{\mathrm P}$, of the propagator is given by
\begin{align}
  &\Pi_{\mathrm P}(k^2) \delta_{\alpha\beta} \nonumber \\
  &= \int \!\! \frac{d^D p}{i(2\pi)^D}
  {\rm tr}\bigl[ \gamma_5 T_\alpha  S^i(p+k/2) \gamma_5 
  T_\beta^\dagger S^j(p-k/2)\bigr],
\label{Pi_P}
\end{align}
where the trace runs over flavor, spinor and color indices. The flavor
$SU(3)$ matrices $T_\alpha$ are: $T_3=\lambda_3$ for $\pi^0$,
$T_6=(\lambda_6+i\lambda_7)/\sqrt{2}$ for ${\mathrm K}^0$ and $T_6^\dagger$
for $\bar{\mathrm K}^0$. Then the self-energy for $\pi^0$, ${\mathrm K}^0$
and $\bar{\mathrm K}^0$ can be written as
\begin{eqnarray}
  \Pi_{\pi}(k^2=m_\pi^2) &=& 2 \Pi_5^{uu}(k^2=m_\pi^2), 
  \label{Pi_pi} \\
  \Pi_{\mathrm K}(k^2=m_{\mathrm K}^2) &=& 2\Pi_5^{su}(k^2=m_{\mathrm K}^2),
  \label{Pi_K}
\end{eqnarray}
where $\Pi_5^{ij}(k^2)$ is defined by
\begin{align}
  \Pi_5^{ij}&(k^2)= \! \int \!\! \frac{d^D p}{i(2\pi)^D}
     \, {\rm tr}\bigl[ \gamma_5 S^i(p+k/2) \gamma_5 
     S^j(p-k/2)\bigr]  \nonumber \\
  &= -\frac12 \Bigl( \frac{\phi_i}{m_i^*} + \frac{\phi_j}{m_j^*} \Bigr)
     +\frac12 \bigl[ k^2 - (m_i^* - m_j^*)^2 \bigr]\,I_{ij},
\label{Pi_5}
\end{align}
with
\begin{align}
  I_{ij}=\int \!\! \frac{d^D p}{i(2\pi)^D} \,
        \frac{{\rm tr}1}{(p^2-m_i^{*\,2})\bigl[(p-k)^2-m_j^{*\,2}\bigr]}.
\label{I_ij}
\end{align}
The integral above becomes 
\begin{align}
  I_{ij} = &\int \!\! \frac{d^{D-1} p}{i(2\pi)^{D-1}}
     \frac{{\rm tr}1}{D_{ij}^{+}} \Biggl\{ \sum_{i\leftrightarrow j}
       \frac{1}{2E_i} \biggl[ 1 - \sum_{s=\pm}
         \frac{S_{ji}}{D_{ij}^{-}} f(E_i^{s}) \biggr] \nonumber \\
    &\,\, -\frac{k_0}{D_{ij}^{-}}
       \left[ f(E_i^{+})-f(E_i^{-})-f(E_j^{+})+f(E_j^{-})\right]
      \Biggr\}
\end{align}
where $D_{ij}^{\pm}=(E_{i}\pm E_{j})^2-k^2$ and $S_{ij} = m_i^{*\,2}-m_j^{*\,2}-k^2$.
In practice we first solve the gap equations to get the chiral condensates
$\phi_i$ and constituent quark masses $m_i^*$, then we scan the solution of
Eq.(\ref{pk_mass}) in evaluating $m_{\pi}$ and $m_{\mathrm K}$. In the actual
numerical calculations, we choose the rest frame $k^{\mu}=(k^0,{\boldsymbol 0})$.

It should be noted that $I_{ij}$ as a function of $k^0$ (${\boldsymbol k}={\boldsymbol 0}$)
has a singularity. Since the pion self-energy $\Pi_{\pi}$
consists of the function $I_{uu}$, it is divergent at the singularity.
$\Pi_{\pi}(k^2)$ has a cut in the region $k^{0} \geq 2|m_u^*|$ on the
real axis. One should move the other Riemann sheet to calculate
$\Pi_{\pi}(k^2)$ at the cut. 
The vacuum part $I_{uu}^{\mathrm (v)}$ is written as
\begin{eqnarray}
  I_{uu}^{\mathrm (v)}=\frac{N_c}{(2\pi)^{D/2}}\Gamma(2-\frac{D}2)\int_0^1\!\!dx 
  L_{uu}^{D/2-2} (k^2) ,
\label{I_ij0}
\end{eqnarray}
where 
\begin{eqnarray}
  L_{uu}(k^2)= m_u^{*2}  - k^2 x(1-x).
\end{eqnarray}
The function $I_{uu}(k^2)$ is divergent at $k^2=4m_u^{*\,2}$ for $D<4$. Then
$\Pi_{\pi}(k^2)$ has the singularity at $k^0 = 2|m_u^{*}|$ which
is confirmed in Fig.~\ref{m_pi_pole} where $D=2.47$ (Case $\chi_{170}$
with $m_u=5.5$MeV). 
The divergence appears at the resonance position. It does not mean the existence
of a bound state except for $k^0=2|m_u^*|$. In Fig.~\ref{m_pi_pole} three
solutions are observed for ${\mathcal F}_\pi=0$ at $T=271$ and 279MeV.
Two of the solutions are located near the singularity at $k^0=2|m_u^*|$. 
We take the solution with the larger $k^0$ to express the pion mass. Thus the 
meson masses jump as functions of $T$ near the critical temperature.

As is shown in Fig.~\ref{m_pi_pole}, the larger solution disappears for 
$T \gtrsim 279$MeV. In this case we determine  the meson mass as $k^0$ at the 
maximum of ${\mathcal F}_\pi$. On the contrary, the reliable solution
always exists and the smaller two solutions are degenerate
above the critical chemical potential at $T=10$MeV.

\subsection{$\pi$ and ${\mathrm K}$ decay constants}
Here, we display the equations determining $f_{\mathrm P}$, the pion and
kaon decay constants, with ${\mathrm P}(=\pi,\,{\mathrm K})$.
\begin{align}
&ik_\mu f_{\mathrm P} \delta_{\alpha\beta} \nonumber \\
&= - M_0^{4-D} \!\int \! \frac{d^D p}{(2\pi)^D}
  {\rm tr}\left[\gamma_\mu \gamma_5 \frac{T_\alpha}{2} 
  S^i g_{\mathrm Pqq}(0) \gamma_5 T_\beta^\dagger S^j \right],
\label{decay_pre}
\end{align}
where the renormalization scale $M_0$ is introduced so that the decay
constant has a correct mass dimension, dim$(f_{\mathrm P})=1$. The trace
is with respect to the flavor, spinor and color indices. The
meson-to-quark-quark coupling, $g_{\mathrm Pqq}$, is defined by
\begin{equation}
g_{\mathrm Pqq}(k^2)^{-2} =  M_0^{4-D}
       \frac{\partial \Pi_{\mathrm P}(k^2)}{\partial k^2} . 
\label{g_Pqq}
\end{equation}
Eq.(\ref{decay_pre}) can be rewritten in the following form
\begin{align}
f_{\mathrm P}^2=-M_0^{4-D}g_{\mathrm Pqq} I_{\mathrm P}^2 / k^2,
\label{decay}
\end{align}
where $I_{\pi}=m_u^*I_{uu}(k^2)k_{\mu}$ and
\begin{align}
  I_{\mathrm K}={\rm tr}1\int \!\! \frac{d^D p}{i(2\pi)^D}\,
    \frac{(m_s^*-m_u^*)p_{\mu}-m_s^*k_{\mu}}
         {(p^2-m_s^{*\,2})\bigl[(p-k)^2-m_u^{*\,2}\bigr]}.\nonumber
\label{decay}
\end{align}
By inserting the values of the constituent quark masses $m_i^*$ and the
renormalization scale $M_0$, one can easily evaluate the decay constants.

\subsection{$\eta$ and $\eta'$ mesons}
The equation which determines $\eta$ meson masses becomes $2\times 2$
matrix due to the well known \ua anomaly. In the leading order of the
$1/N_c$ expansion, the propagator of the $\eta-\eta^{\prime}$ system
becomes
\begin{equation}
  {\bf \Delta}^{\!\! +}(k^2) = 2{\bf K}^{\! +} 
       [1-2{\bf K}^{\! +} {\bf \Pi}(k^2)]^{-1} ,
\label{pro^+}
\end{equation}
where ${\bf K}^{\! +}$ and ${\bf \Pi}$ are given by
\begin{eqnarray}
  {\bf K}^{\! +} &=& \left(
\begin{array}{cc}
  K_{00} & K_{08} \\
  K_{80} & K_{88}
\end{array}
  \right) ,
\label{K^+} \\
  {\bf \Pi} &=& \left(
\begin{array}{cc}
  \Pi_{00} & \Pi_{08} \\
  \Pi_{80} & \Pi_{88}
\end{array}
  \right) ,
\label{Pi}
\end{eqnarray}
with
\begin{align*}
  &K_{00}=G + \frac{1}{3} K(\phi_s + 2 \phi_u), \\ 
  &K_{88}=G + \frac{1}{6} K(\phi_s - 4 \phi_u), \\ 
  &K_{08}=K_{80}=\frac{\sqrt{2}}{6}K(\phi_s-\phi_u), 
\end{align*}
and
\begin{align*}
  &\Pi_{00}(k^2)=\frac{2}{3}\left[ 2\Pi_5^{uu}(k^2)+\Pi_5^{ss}(k^2) \right],
  \\ 
  &\Pi_{88}(k^2)=\frac{2}{3}\left[ \Pi_5^{uu}(k^2)+2\Pi_5^{ss}(k^2) \right],
  \\ 
  &\Pi_{08}(k^2)=\Pi_{80}(k^2)=\frac{2\sqrt{2}}{3} \left[ \Pi_5^{uu}(k^2)
                 -\Pi_5^{ss}(k^2) \right].
\end{align*}
The condition which determines the $\eta$ and $\eta^{\prime}$ masses is,
\begin{equation}
\mbox{det}[ 1-2{\bf K}^{\! +} {\bf \Pi}(x^2)]=0.
\label{eta}
\end{equation}
Using the constituent quark masses derived from the gap equations, we
search for the appropriate solutions for $m_{\eta}$ and 
$m_{\eta^{\prime}}$.

\subsection{Topological susceptibility}
The topological susceptibility $\chi$ is given by the correlation 
function between the topological charge densities $Q(x)$ at different
points,
\begin{equation}
 \chi=\,\int\! d^4x\langle 0| TQ(x)Q(0)|0\rangle_{\rm connected}\,\,,
\end{equation}
here the charge density is defined as
\begin{align}
Q(x)\equiv\frac{g^2}{32\pi^2}F^a_{\mu\nu} \tilde{F}^{a\mu\nu}
= 2K\, {\rm Im} [\det \bar{q}(1-\gamma_5)q],
\end{align}
with $g$, the strong coupling constant of QCD, and $F^a_{\mu\nu}$, the
gluon field strength. The explicit form of $\chi$ is given below~%
\cite{Fukushima:2001prc},
\begin{align}
 \chi
 =\, -&\frac{4K^2}{M_0^{D-4}} \phi_u^2 
 \left[ \phi_u \phi_s  
 \left(\frac{2\phi_s}{m_u^*} + 
 \frac{\phi_u}{m_s^*} \right) \right.
\nonumber \\
 &-\left\{ \frac{1}{\sqrt{6}} (2\phi_s + \phi_u)
 \bigl( \Pi_{00}(0), \Pi_{08}(0) \bigr) \right.
\nonumber \\
 &\quad \,\,\, +\left. \frac{1}{\sqrt{3}} (\phi_s - \phi_u)
 \bigl( \Pi_{08}(0), \Pi_{88}(0) \bigr)
 \right\} \Delta^+(0)
\nonumber \\
 &\times\left\{ \frac{1}{\sqrt{6}} (2\phi_s + \phi_u)
 \left(
\begin{array}{c}
 \Pi_{00}(0) \\
 \Pi_{08}(0) 
\end{array}\right) \right.
 \nonumber \\
 &\quad \,\,\, +\left. \frac{1}{\sqrt{3}} (\phi_s - \phi_u)
 \left( \left.
\begin{array}{c}
 \Pi_{08}(0) \\
 \Pi_{88}(0)
\end{array} \right) \right\} \right] .
\label{chi}
\end{align}
After solving the gap equations, one can numerically calculate
this quantity by inserting the values of $\phi_i$ and $m_i^*$.

\section{Parameter sets}
\label{app_para}
Here we present the parameter sets obtained in~\cite{Inagaki:2010nb}.
In the Case $m_{\eta}$ and $\chi_{179}$, there exist two sets of
parameters, and we distinguish between these cases by using the
superscript LD (lower dimension) and HD (higher dimension).
\begin{table}
 \caption{Case $m_{\eta}^{\rm LD}$.}
 \label{para_etaLD}
 \begin{ruledtabular}
 \begin{tabular}{lccccc}
$m_u$ & $m_s$ & $G$ & $K$ & $M_0$ & $D$ \\
\hline %
$3.0$ & $84.9$ & $-0.0195$ & $9.02 \times 10^{-7}$ & $118$ & $2.29$ \\
$4.0$ & $118$ & $-0.0174$ & $9.17 \times 10^{-7}$ & $113$ & $2.38$ \\
$5.0$ & $156$ & $-0.0162$ & $9.49 \times 10^{-7}$ & $108$ & $2.47$ 
 \end{tabular}
 \end{ruledtabular}
 \caption{Case $m_{\eta}$.}
 \label{para_eta}
 \begin{ruledtabular}
 \begin{tabular}{lccccc}
$m_u$ & $m_s$ & $G$ & $K$ & $M_0$ & $D$ \\
\hline %
$3.0$ & $79.0$ & $-0.0130$ & $2.29 \times 10^{-7}$ & $107$ & $2.37$ \\
$4.0$ & $106$ & $-0.00748$ & $8.26 \times 10^{-8}$ & $92.0$ & $2.52$ \\
$5.0$ & $134$ & $-0.00357$ & $1.99 \times 10^{-8}$ & $73.2$ & $2.69$ \\
$5.5$ & $147$ & $-0.00231$ & $8.40 \times 10^{-9}$ & $62.4$ & $2.77$ \\
$6.0$ & $162$ & $-0.00142$ & $3.23 \times 10^{-9}$ & $50.9$ & $2.87$
 \end{tabular}
 \end{ruledtabular}
\end{table}
\begin{table}
 \caption{Case $\chi_{170}$.}
 \label{para_chi170}
 \begin{ruledtabular}
 \begin{tabular}{lccccc}
$m_u$ & $m_s$ & $G$ & $K$ & $M_0$ & $D$ \\
\hline %
$3.0$ & $77.1$ & $-0.0168$ & $2.23 \times 10^{-7}$ & $120$ & $2.28$ \\
$4.0$ & $106$ & $-0.0143$ & $2.11 \times 10^{-7}$ & $116$ & $2.36$ \\
$5.0$ & $134$ & $-0.0119$ & $1.80 \times 10^{-7}$ & $112$ & $2.43$ \\
$5.5$ & $150$ & $-0.0109$ & $1.62 \times 10^{-7}$ & $110$ & $2.47$ \\
$6.0$ & $166$ & $-0.00992$ & $1.48 \times 10^{-7}$ & $109$ & $2.50$
 \end{tabular}
 \end{ruledtabular}
\end{table}
\begin{table}
 \caption{Case $\chi_{179}$}
 \label{para_chi179}
 \begin{ruledtabular}
 \begin{tabular}{lcccccc}
$m_u$ & $m_s$ & $G$ & $K$ & $M_0$ & $D$ \\
 \hline %
$3.0$ & $78.0$ & $-0.0170$ & $2.70 \times 10^{-7}$ & $120$ & $2.28$ \\
$4.0$ & $106$ & $-0.0144$ & $2.52 \times 10^{-7}$ & $115$ & $2.36$ \\
$5.0$ & $136$ & $-0.0120$ & $2.11 \times 10^{-7}$ & $111$ & $2.44$ \\
$5.5$ & $152$ & $-0.0109$ & $1.90 \times 10^{-7}$ & $110$ & $2.47$ \\
$6.0$ & $168$ & $-0.00995$ & $1.70 \times 10^{-7}$ & $108$ & $2.50$
 \end{tabular}
 \end{ruledtabular}
 \caption{Case $\chi_{179}^{\rm HD}$}
 \label{para_chi179HD}
 \begin{ruledtabular}
 \begin{tabular}{lcccccc}
 $m_u$ & $m_s$ & $G$ & $K$ & $M_0$ & $D$ \\
 \hline %
 $3.0$ & $74.8$ & $-4.24 \times 10^{-5}$ & $1.38 \times 10^{-13}$ & $2.91$ & $3.22$ \\
 $4.0$ & $100$ & $-6.91 \times 10^{-8}$ & $3.60 \times 10^{-19}$ 
       & $1.15 \times 10^{-25}$ & $3.90$
 \end{tabular}
 \end{ruledtabular}
\end{table}

\begin{table}
 \caption{Case Cutoff.}
 \label{para_cut}
 \begin{ruledtabular}
 \begin{tabular}{lcccc}
$m_u$  & $m_s$  & $G\Lambda^2$ & $K\Lambda^5$ & $\Lambda$ \\
\hline %
$3.0$  & $89.5$ & $1.55$ & $8.34$ & $960$ \\
$4.0$  & $110$  & $1.60$ & $8.38$ & $797$ \\
$5.0$  & $128$  & $1.71$ & $8.77$ & $682$ \\
$5.5$  & $136$  & $1.81$ & $9.17$ & $630$ \\
$5.87$ & $139$  & $2.09$ & $10.1$ & $580$
 \end{tabular}
 \end{ruledtabular}
\end{table}

In Tab.~\ref{para_cut}, we also present the parameter sets of the Case
Cutoff. In the cutoff regularization, the number of parameters is five, and
these are fixed by using the four physical observables $m_{\pi}$, $f_{\pi}$,
$m_{\mathrm K}$, $m_{\eta^{\prime}}$ for several values of $m_u$.


\end{document}